\documentclass[aip,jcp,11pt,citeautoscript,amsmath,amssymb,reprint]{revtex4-1}

\newif\ifpdf\ifx\pdfoutput\undefined\pdffalse\else\pdfoutput=1\pdftrue\fi

\usepackage{graphicx}
\usepackage[usenames]{color} 
\usepackage{comment} 
\usepackage{bm}
\newcounter{abc}

\newcommand{\be}{\begin{equation}} 
\newcommand{\ee}{\end{equation}}
\newcommand{\bea}{\begin{eqnarray}} 
\newcommand{\eea}{\end{eqnarray}}

\newcommand{\rv}{\bm{r}} \newcommand{\f}{f} 
 
\newcommand{\p}{^{(0)}} 

\newcommand{\effe}{{\mathcal{E}}}

\begin{document}
\title{\bf Polydispersity induced solid-solid transitions in model colloids}
\vspace{0.6cm}

\author{Peter Sollich} \affiliation{King's College London, Department of
Mathematics, Strand, London WC2R 2LS, United Kingdom.} 
\author{Nigel B. Wilding} \affiliation{Department of Physics, University of Bath, Bath BA2
7AY, United Kingdom.}

\begin{abstract}
Specialized Monte Carlo simulation techniques and moment free energy
method calculations, capable of treating fractionation exactly, are 
deployed to study the crystalline phase behaviour of an assembly of
spherical particles described by a top-hat ``parent'' distribution of
particle sizes. An increase in either the overall density or the degree of
polydispersity is shown to generate a succession of phase transitions in
which the system demixes into an ever greater number of face-centred
cubic ``daughter'' phases. Each of these phases is strongly
fractionated: it contains a much narrower distribution of particle sizes
than is present in the system overall. Certain of the demixing
transitions are found to be nearly continuous, accompanied by
fluctuations in local particle size correlated over many lattice
spacings. We explore possible factors controlling the stability of the
phases and the character of the demixing transitions.
\end{abstract}
\maketitle
\section{Introduction} 
\label{sec:intro}

Hard spherical particles can be packed to fill maximally just over
$74\%$ of space, in the face centred cubic (fcc) structure
\cite{Hales2006}. For systems in thermal equilibrium such as a colloidal
suspension, this structure remains preferred
~\cite{Woodcock1997a,Bruce1997a} for packing fractions down to about
$55\%$~\cite{Alder1957} where melting occurs. But what is the
thermodynamically optimal structure for spherical colloids that are
``polydisperse'', i.e.\ have a spread of diameters? Polydispersity should
act to destabilize a colloidal crystal because of the difficulty of
accommodating a range of particle sizes within a single lattice
structure; but despite sustained attention spanning over three decades
(see
e.g.~\cite{Dickinson1978,Barrat1986,Bartlett1998,Sear1998,Phan1998,Lacks1999,Chaudhuri2005,Fernandez2007,Yang2009}),
there is a lack of consensus as to what stable structures arise instead.

Attempts to address this matter have focused on the use of analytical
theory and simulation to predict the fate of a single crystal in the
dense regime (above typical fluid densities) when the degree of
polydispersity becomes large. Broadly speaking, two incompatible
proposals have emerged: either the system demixes into multiple
coexisting crystalline phases \cite{Bartlett1998,Sear1998} or, 
alternatively, crystalline phases disappear altogether \cite{Lacks1999},
the crystal being replaced by an {\em equilibrium} glassy phase
\cite{Chaudhuri2005}.  Ideally, of course, one should like to settle the
matter as to which (if either) of these scenarios is correct by simply
performing an experiment with a suitable suspension of
colloids. But the inhibition of diffusion in crystalline phases is
expected to render solid-solid demixing transitions largely unobservable on
experimental timescales, even if they are thermodynamically favoured\footnote{Though see reference \cite{Byelov2010} for a recent
experimental observation of solid-solid phase separation in polydisperse
platelike particles.}. This would -- on the face of it -- appear to
render our central question moot. However, one should recognize that
even when equilibrium is itself unattainable in practical situations,
independent knowledge of the stable state represents an important
baseline for interpreting dynamical properties of colloidal systems
(such as crystallization kinetics \cite{Auer2001a}) which can be
understood in terms of the topology of the free energy surface
\cite{Poon2002}. There are also suggestions \cite{ZacValSanPooCat09}
that the equilibrium phase diagram sheds light e.g.\ on the ability of
a glassy phase to crystallize. 
The question as to the nature of the true stable state
is thus of more than merely academic interest.

In our view, the disparity in the predictions of previous theoretical
and computational work is traceable to the fact that when considering
phase separation, little or no account was taken of ``fractionation'',
i.e.\ the phenomenon whereby the distribution of the particle diameters,
$\sigma$, can vary from one coexisting phase to another
\cite{evans1998,Erne2005}. Indeed it is now well established that
fractionation can radically alter the qualitative features of phase
behaviour in polydisperse systems compared to their monodisperse
counterparts (see \cite{Sollich2001} for a review). Accordingly, it is
essential to fully incorporate its effects if one hopes to describe the
equilibrium phase behaviour of polydisperse systems correctly
\cite{Wilding2010b}.

To quantify fractionation \cite{Salacuse1982} one simply counts, for a certain phase
(labeled $\alpha$), the number density of particles having diameters in
the range $\sigma\ldots\sigma+d\sigma$. This serves
to define a density distribution $\rho^{(\alpha)}(\sigma)$. However, in
real colloidal suspensions,  one has the constraint that the overall
distribution of sizes (across all phases) has a form fixed by the
synthesis of the suspension. This gives rise to a generalized lever
rule:    

\be
\rho^{(0)}(\sigma)=\sum_\alpha\lambda^{(\alpha)}\rho^{(\alpha)}(\sigma),
 \label{eq:lever} 
\ee    
with $\rho^{(0)}(\sigma)$ the ``parent'' density distribution, $\rho^{(\alpha)}(\sigma)$ the ``daughter''
distributions, and  $\lambda^{(\alpha)}$ the fractional volume occupied by phase
$\alpha$ (so that $\sum_\alpha \lambda^{(\alpha)}=1$). Since
the form of the parent is fixed, only its scale is free to vary, e.g.\
by dilution with solvent, and one writes $\rho^{(0)}(\sigma)=n\p
f(\sigma)$, where $n\p$ is the total number density and $f(\sigma)$ is a
prescribed normalized shape function.  The degree of polydispersity,
$\delta$, is then defined as the standard deviation of the parent
distribution $f(\sigma)$, expressed in units of its mean. 

Fractionation greatly complicates the task of determining the phase
behaviour of polydisperse systems compared to their monodisperse
counterparts. To illustrate this, consider (for a given colloidal system)
increasing $n\p$ from an initially low value, i.e.\ following a ``dilution
line'' through the phase diagram of the system. For sufficiently large
$n\p$ the system typically encounters a coexistence region of the phase
diagram, which is entered at a ``cloud'' point \cite{Sollich2001} value of
$n\p$. At and beyond this density the system separates into differently
fractionated daughter phases. However, as a consequence of fractionation
both the daughter distributions $\rho^{(\alpha)}(\sigma)$ themselves and their
associated fractional volumes $\lambda^{(\alpha)}$ depend {\em
non-linearly} on $n\p$. Thus in order to quantify the phase behaviour
one is faced with the challenge of determining the daughter phase
properties for {\em all} values of $n\p$ within the coexistence region
-- a situation which contrasts with the monodisperse case where the
coexistence densities are independent of the total density, whilst the
fractional volumes  depend linearly on it.

One theoretical technique that does take fractionation into account
exactly (within the context of a mean field framework) is the moment
free energy (MFE) method. Previous work using this method by one of us
\cite{Fasolo2004} predicts that, for polydisperse spheres, increasing
$\delta$ or $n\p$ within the 
solid region leads to a succession of phase transitions in which the
system demixes into an ever greater number of differently-fractionated 
daughter phases. Each daughter phase contains a narrower
distribution of particle diameters than the parent. This MFE
calculation thus provides clear evidence for the scenario of multiple
coexisting solids. But it uses approximate free energy expressions,
which for solids 
are derived from those of binary mixtures and implicitly already assume
that all solids are fcc. Independent confirmation of its predictions is
then highly desirable, but has hitherto been lacking. The purpose of the
present report is therefore to provide a definite answer to the question
of the nature of the equilibrium phase behaviour via state-of-the-art
Monte Carlo (MC) simulations, and to compare with MFE calculations;  both
fully provide for fractionation and employ a fixed parent size
distribution. 

Our paper is organized as follows. In Sec.~\ref{sec:models} we introduce
our model systems: size disperse hard spheres (which we have studied by
the MFE method), and soft spheres (which we have studied by MC
simulation). Sec.~\ref{sec:methods} provides a brief description of both
the MFE method and the bespoke MC techniques required for dealing with
fixed polydispersity and fractionation. Thereafter, in
Sec.~\ref{sec:phasediagram}, we report our observations concerning the
phase behaviour of these models, the central finding being that the
original MFE calculations are indeed correct: as $\delta$ and/or $n\p$
are increased, a succession of transitions occurs in which the system
demixes into first two, then three, then four fractionated coexisting
fcc crystalline phases. We analyse the observations to arrive at a
qualitative picture of when a crystalline phase will become unstable to
phase separation. In Sec.~\ref{sec:criticality}, we investigate in
detail the character of these phase transitions, finding that some are
strongly first order, whilst others are quasi-continuous. To quantify
the differences, we introduce and measure a susceptibility that probes
particle size fluctuations. The associated correlation length at the
near continuous transitions is found to extend over several crystal
lattice spacings. Prompted by these observations, we use MFE
calculations to study in detail how the shape of the size distribution
affects the tendency of a given solid phase to exhibit a near continuous
demixing transition. This leads us to a simple approximate criterion that
quantifies this tendency. Finally, in Sec.~\ref{sec:discussion} we
summarize and discuss the significance of our results, and indicate some
issues for future work.

\section{Models}
\label{sec:models}

The systems that we shall consider in this work are assemblies of
spheres interacting either by a repulsive soft sphere potential (as considered by
simulation) or a hard sphere potential (as studied in our MFE
calculations). The soft sphere interaction potential
between two particles $i$ and $j$ with position vectors $\rv_i$ and
$\rv_j$ and diameters $\sigma_i$ and $\sigma_j$ is given by  
\begin{equation} v(r_{ij})=\epsilon(\sigma_{ij}/r_{ij})^{12}\:,
\label{eq:softspheres} 
\end{equation} 
with particle separation $r_{ij}=|\rv_i-\rv_j|$ and interaction radius
$\sigma_{ij}=(\sigma_i+\sigma_j)/2$. The choice of this potential rather
than hard spheres is made on pragmatic grounds; in our isobaric SGCE
simulations (to be reported below), any MC contraction of the simulation box that leads to an
infinitesimal overlap of two hard spheres would always be rejected, so
(particularly at high densities) we can expect higher MC acceptance
rates using a ``softer'' potential. In common with hard spheres, the
monodisperse version of our model freezes into an fcc crystalline
structure \cite{Hansen1970,Hoover1970,Wilding2009a}, and temperature
only plays the role of a scale: the thermodynamic state depends not on
$n\p$ and $T$ separately but only on the combination
$n\p(\epsilon/k_{\mathrm{B}}T)^{1/4}$. Phase diagrams for different $T$
then scale exactly onto one another, and we can fix
$\epsilon/k_{\mathrm{B}}T=1$.

In all cases we consider parent size distributions of the top-hat form:
\begin{equation}
f(\sigma)=\left\{
\begin{array}{ll}
(2c)^{-1} & \mbox { if $1-c\le \sigma/\bar{\sigma} \le 1+c$} \\
~~0      &  \mbox { otherwise }
\end{array}
\right. .
\label{eq:th}
\end{equation}
Here the width parameter $c$ controls the
degree of polydispersity $\delta=c/\sqrt{3}$. In the following we use the mean particle diameter $\bar{\sigma}$ as our unit of length.

\section{Methodologies} 
\label{sec:methods}

\subsection{Analytical calculations: the moment free energy method}
\label{sec:MFE}

\newcommand{\fex}{f^{\rm ex}}

Calculating analytically the phase behaviour of polydisperse systems
is a challenging problem~\cite{Sollich2002}. This is because for each
of the infinitely many different particle sizes $\sigma$ one has a
separate conserved density $\rho(\sigma)$. Effectively one thus has to
study the thermodynamics of an infinite mixture, where e.g.\ from
the Gibbs rule there is no upper limit on the number of phases that
can occur.

The moment free energy (MFE)
method~\cite{SolWarCat01,Warren98,SolCat98,Sollich2002} is designed to
get around this issue by effectively projecting the infinite
mixture problem down to that for a finite mixture of
``quasi-species''. This is possible when the free energy density has a
so-called truncatable form,
\begin{equation}
f = k_{\mathrm{B}}T \int d\sigma \rho(\sigma) \left[\ln(\rho(\sigma))-1\right] +
\fex(\{\rho_i\})\ ,
\label{f:general}
\end{equation}
where the excess part $\fex$ depends on a finite number of moments of
the density distribution,
\begin{equation}
\rho_i = \int d\sigma \rho(\sigma) w_i(\sigma)\ .
\end{equation}
This truncatable structure obtains for a large number of models of
mean field type. Importantly for our purposes, it is also found in
accurate free energy expressions for polydisperse hard spheres, with
the simple weight functions $w_i(\sigma)=\sigma^i$ ($i=0,1,2,3$).
Specifically, we use the free energy developed by Bartlett~\cite{Bartlett97} on the
basis of the simulation data of Kranendonk {\em et
al.}~\cite{KraFre91} for binary mixtures. As mentioned above, this
effectively presupposes fcc structures for all solids, so that
validation e.g.\ by simulations, as provided in this paper, is important.

The MFE method provides a way of expressing the ideal contribution to the
free energy from Eq.~(\ref{f:general}), which depends on the complete
shape of the density distribution, in terms of the moment densities
$\rho_i$. The result is the moment free energy. The key
feature of the method is that if one then treats the quasi-species
densities $\rho_i$ as if they were densities of ordinary particle
species, and calculates phase equilibria accordingly, the results for
cloud points are fully exact. Within coexistence regions, the method
can be extended by including additional moments~\cite{SpeSol02}. Their weight
functions can be chosen adaptively, and using the resulting
approximation as an initialization~\cite{SpeSol03a}, the exact phase equilibrium conditions can
then be solved numerically, even if e.g.\ three or four daughter phases are
present.
Overall, the MFE approach is therefore
the method of choice for our current investigation. We do not give
further details of the numerical implementation here as these are set
out in full in Ref.~\cite{Fasolo2004}.

\subsection{Simulation: phase behaviour within the isobaric semi-grand canonical ensemble}
\label{sec:}

The appropriate ensemble for determining phase behaviour in dense
assemblies of polydisperse particles is the isobaric variant of the
semi-grand canonical ensemble (SGCE) \cite{Wilding2010b}.  Within this
ensemble, the particle number $N$, pressure $p$, temperature $T$, {\em
and} a distribution of chemical potential differences
$\tilde\mu(\sigma)$ are all prescribed, while the system volume
$V$, the energy, and the form of the instantaneous density distribution
$\rho(\sigma)$ all fluctuate \cite{Kofke1988}.  The fluctuations in
$\rho(\sigma)$ are linked to the volume fluctuations by the
relation 
$V\int\rho(\sigma)d\sigma=N$. Importantly, they permit the sampling of many
realizations of the polydisperse disorder, thus ameliorating finite-size
effects. Moreover, in conjunction with volume fluctuations,  they
facilitate separation into differently fractionated phases. Coexistence
of two or more phases is signalled by a multimodal form for the
distribution of some order parameter such as the overall number density
$n=N/V$ or the volume fraction $\eta$, which for a phase with density
distribution $\rho(\sigma)$ can be written as $\eta =\int d\sigma \rho(\sigma)
(\pi/6)\sigma^3$.
 
Operationally, the sole difference between the isobaric
semi-grand canonical ensemble and the constant-{\em NpT} ensemble
\cite{frenkelsmit2002} is that one implements MC updates that select a
particle at random and attempt to change its diameter $\sigma$ to
$\sigma'$ by a
random amount $\sigma'-\sigma$ drawn from a zero-mean uniform
distribution. This proposal 
is accepted or rejected with a Metropolis probability controlled by the
change in the internal energy and chemical potential \cite{Kofke1988}:
\[
p_{\rm acc}={\rm min}\left[1,\exp{(-\beta[\Delta \Phi+\tilde\mu(\sigma)-\tilde\mu(\sigma^\prime)])}\right]\:,
\]
where $\Delta \Phi$ is the internal energy change associated with the
resizing operation and $\beta=1/(k_{\mathrm{B}}T)$.

For SGCE simulations of a polydisperse system at some given $N$ and $T$,
it is necessary to first determine the pressure $p$ and distribution of
chemical potential differences $\tilde\mu(\sigma)$ such that a suitably
defined ensemble-averaged density distribution matches the prescribed
parent $\rho\p(\sigma)=n\p f(\sigma)$. Unfortunately, this task is
complicated by the fact $p$ and $\tilde\mu(\sigma)$ are unknown
{\em functionals} of the parent~\cite{Wilding2003a}. To solve this
problem -- and hence determine correct coexistence properties -- we shall
employ a version of a scheme originally proposed in the context of grand
canonical ensemble studies of polydisperse phase coexistence
\cite{Buzzacchi2006} and later extended to the
SGCE~\cite{Wilding2009,Wilding2010b}, the latter implementation of which
we now summarize. 

The strategy is as follows. For a given choice of $n\p$ and temperature
$T$, one tunes $p$, $\tilde\mu(\sigma)$ and the $\lambda^{(\alpha)}$ iteratively
within a histogram reweighting (HR) framework \cite{ferrenberg1989}, such
as to simultaneously satisfy both a generalized lever rule {\em and}
equality of the probabilities of occurrence of the phases, i.e.
\setcounter{abc}{1} 
\bea 
\label{eq:methoda} 
n\p\f(\sigma) &=&\sum_\alpha \lambda^{(\alpha)}\rho^{(\alpha)}(\sigma),\\ \addtocounter{abc}{1}
\addtocounter{equation}{-1} 
\effe &=&0\:,
\label{eq:methodb} 
\eea 
\setcounter{abc}{0}%
with $\effe$ as defined in Eq.~(\ref{effe_def}) below.
In the first of these constraints, Eq.~(\ref{eq:methoda}), the ensemble
averaged daughter density distributions $\rho^{(\alpha)}(\sigma)$ are
assigned by averaging only over configurations belonging to the
respective phase, distinguishable via the multimodal character of the
order parameter distribution $p(n)$. The deviation of the weighted sum of the daughter
distributions $\bar\rho(\sigma)\equiv \sum_\alpha \lambda^{(\alpha)}\rho^{(\alpha)}(\sigma)$ from the target $n\p\f(\sigma)$ is
conveniently quantified by a ``cost'' value:
\begin{equation}  \Delta\equiv\int
\mid\bar\rho(\sigma)-n\p\f(\sigma)\mid d\sigma \;.  
\label{eq:costfn} 
\end{equation}  
In the second constraint, Eq.~(\ref{eq:methodb}),
\begin{equation}
\effe\equiv \sum_\alpha \left( p^{(\alpha)}-\frac{1}{m} \right)^2 
\label{effe_def}
\end{equation}
provides a measure of the extent to which the probability of
each phase occuring, $p^{(\alpha)}$, is equal for each of the $m$
coexisting phases. Imposing this equality ensures that finite-size errors
in coexistence parameters are exponentially small in the system
volume~\cite{Borgs1992,Buzzacchi2006}.

The iterative determination of $p, \tilde\mu(\sigma)$ and $\lambda^{(\alpha)}$ such as to
satisfy Eqs.~(\ref{eq:methoda}) and (\ref{eq:methodb}) proceeds thus:

\begin{enumerate}

\item Guess initial values of the fractional volumes $\lambda^{(\alpha)}$
corresponding to the chosen value of $n\p$. Usually if one starts near a
cloud point, the fractional volume of the incipient phase will be close
to zero.

\item Tune the pressure $p$ (within the HR scheme) such as to minimize
$\Delta$.

\item Similarly tune $\tilde\mu(\sigma)$ (within the HR scheme)
such as to minimize $\Delta$.

\item Measure the corresponding value of $\effe$.

\item If $\effe< {\rm tolerance}$, finish, otherwise vary
$\lambda^{(\alpha)}$ (within the HR scheme) and repeat
from step 2.

\end{enumerate}

In step $3$
the minimization of $\Delta$ with respect to variations in
$\tilde\mu(\sigma)$ is most readily achieved~\cite{Wilding2002d} using the
following simple iterative scheme for $\tilde\mu(\sigma)$:
\begin{equation}
\beta\tilde\mu_{k+1}(\sigma)=\beta\tilde\mu_k(\sigma)+a\ln\left( \frac{n\p
f(\sigma)} {\bar\rho(\sigma)}\right)\;, 
\label{eq:update} 
\end{equation}
for iteration $k\to k+1$. This update is applied simultaneously to all
entries in the histogram of $\tilde\mu(\sigma)$, and thereafter the
distribution is shifted so that $\tilde\mu(\sigma_0)=0$, where
$\sigma_0$ is the chosen reference size. The quantity $0<a<1$
appearing in Eq.~(\ref{eq:update}) is a damping factor, the value of
which may be tuned to optimize the rate of convergence. Note that (as
described in \cite{Buzzacchi2006}) it is important that one minimizes
$\Delta$ and $\effe$ to a very high precision in order to ensure that
the finite-size effects are exponentially small in the system size.
Typically we iterated until both were less than $10^{-12}$.

The values of $\lambda^{(\alpha)}$ and $p$ resulting from the
application of the above procedure are the desired fractional volumes
and pressure corresponding to the nominated value of $n\p$. As
mentioned above, daughter phase densities and
volume fractions are obtainable by monitoring the multimodal nature of the
order parameter distribution $p(n)$, which allows configurational
properties to be assigned to a given phase~\cite{Wilding2009}.

\section{Phase diagram and solid stability}
\label{sec:phasediagram}

\begin{figure}[h]
\begin{center}
\includegraphics[type=pdf,ext=.pdf,read=.pdf,width=0.85\columnwidth,clip=true]{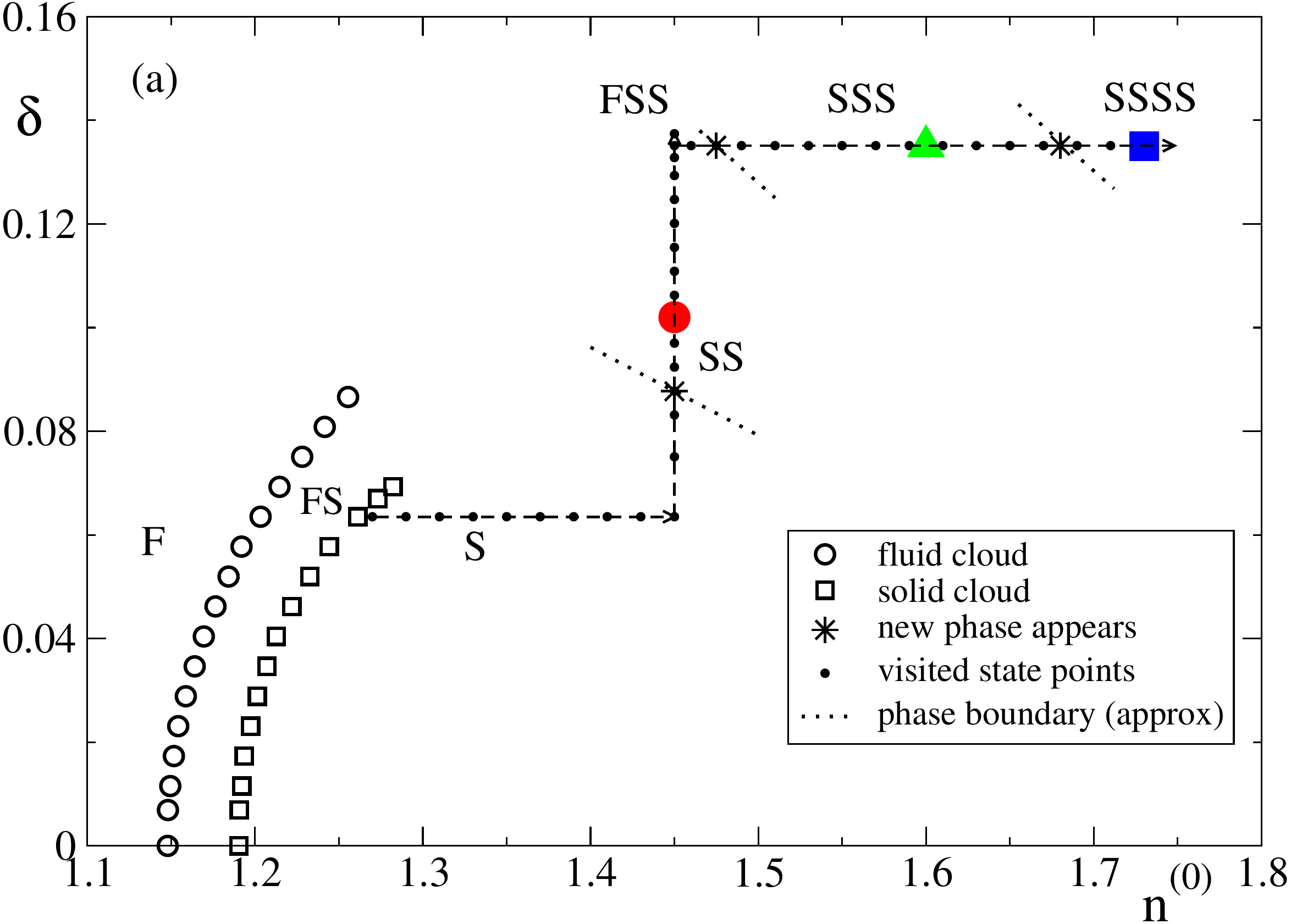}\\
\hspace*{-1.2mm}\includegraphics[type=pdf,ext=.pdf,read=.pdf,width=0.84\columnwidth,clip=true]{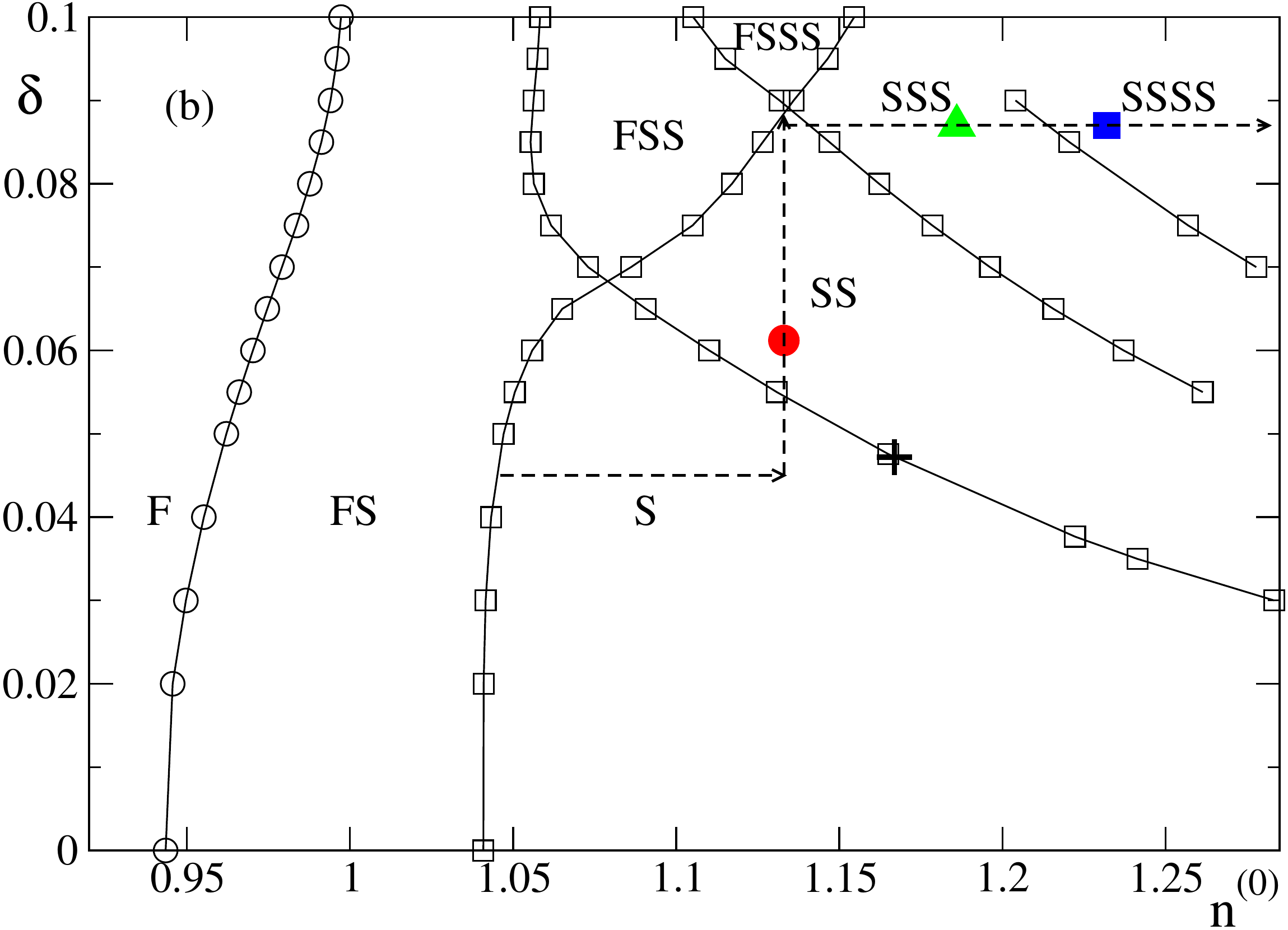}
\end{center}
\caption{(Color online). {\bf (a)} Simulation results for the partial
phase diagram of the
model~(\ref{eq:softspheres}) with parent distribution~(\ref{eq:th}).
Asterisks: points where new solid phases appear; dashed lines: phase
boundary slopes found by histogram reweighting.
F=fluid, S=solid.  
Colored symbols: state points considered in
Fig.~\protect\ref{fig:daughters2}, \protect\ref{fig:daughters3} and \protect\ref{fig:daughters4}.
{\bf (b)} MFE calculation of phase diagram of hard spheres with the same
parent form. The dashed line shows a trajectory comparable to that
followed by the simulations. The cross marks the critical point for
the S--SS transition.}
\label{fig:part_pd} \end{figure}

\begin{figure}[t]
\begin{center}
\includegraphics[type=pdf,ext=.pdf,read=.pdf,width=0.85\columnwidth,clip=true]{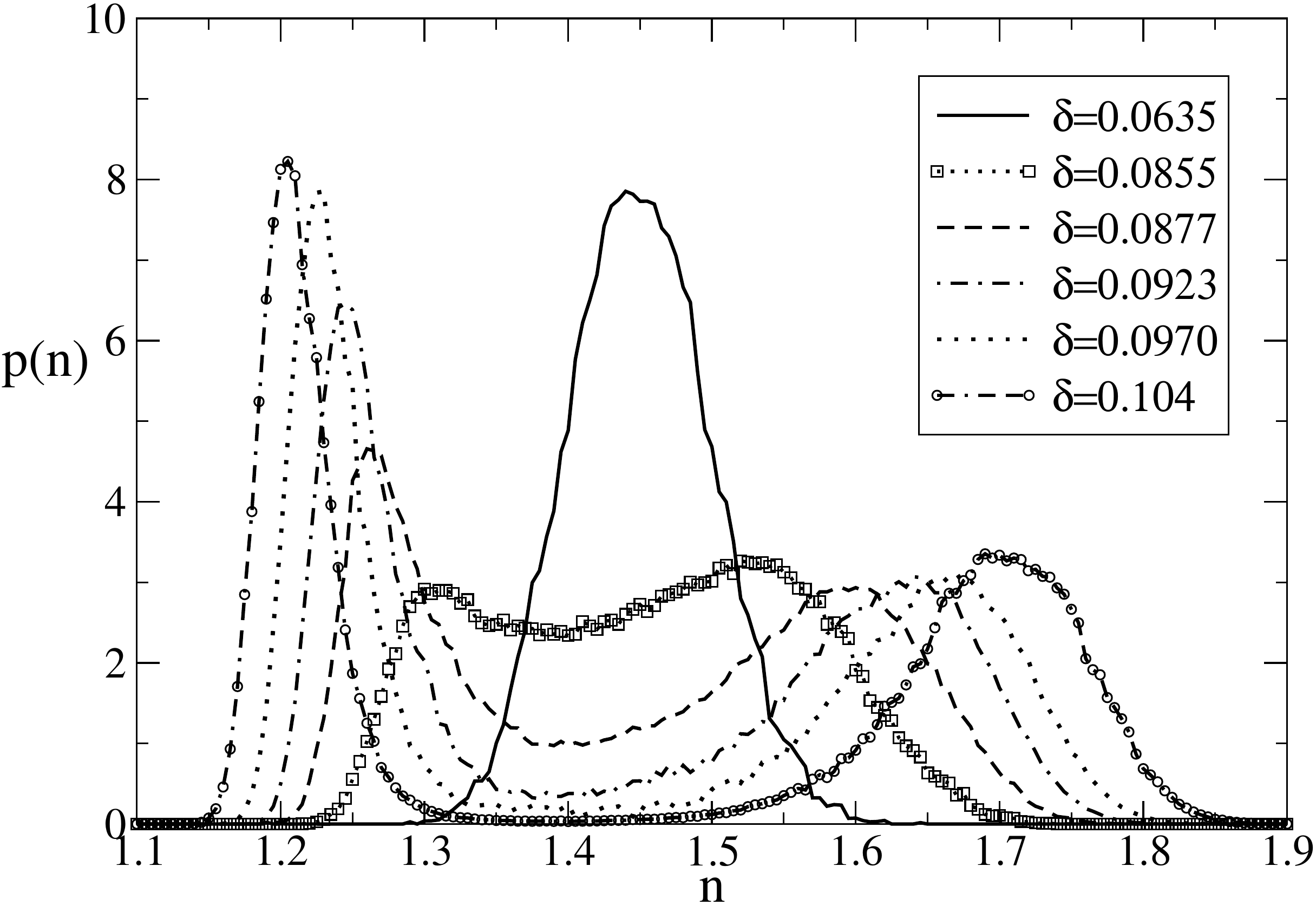}\\
\includegraphics[type=pdf,ext=.pdf,read=.pdf,width=0.85\columnwidth,clip=true]{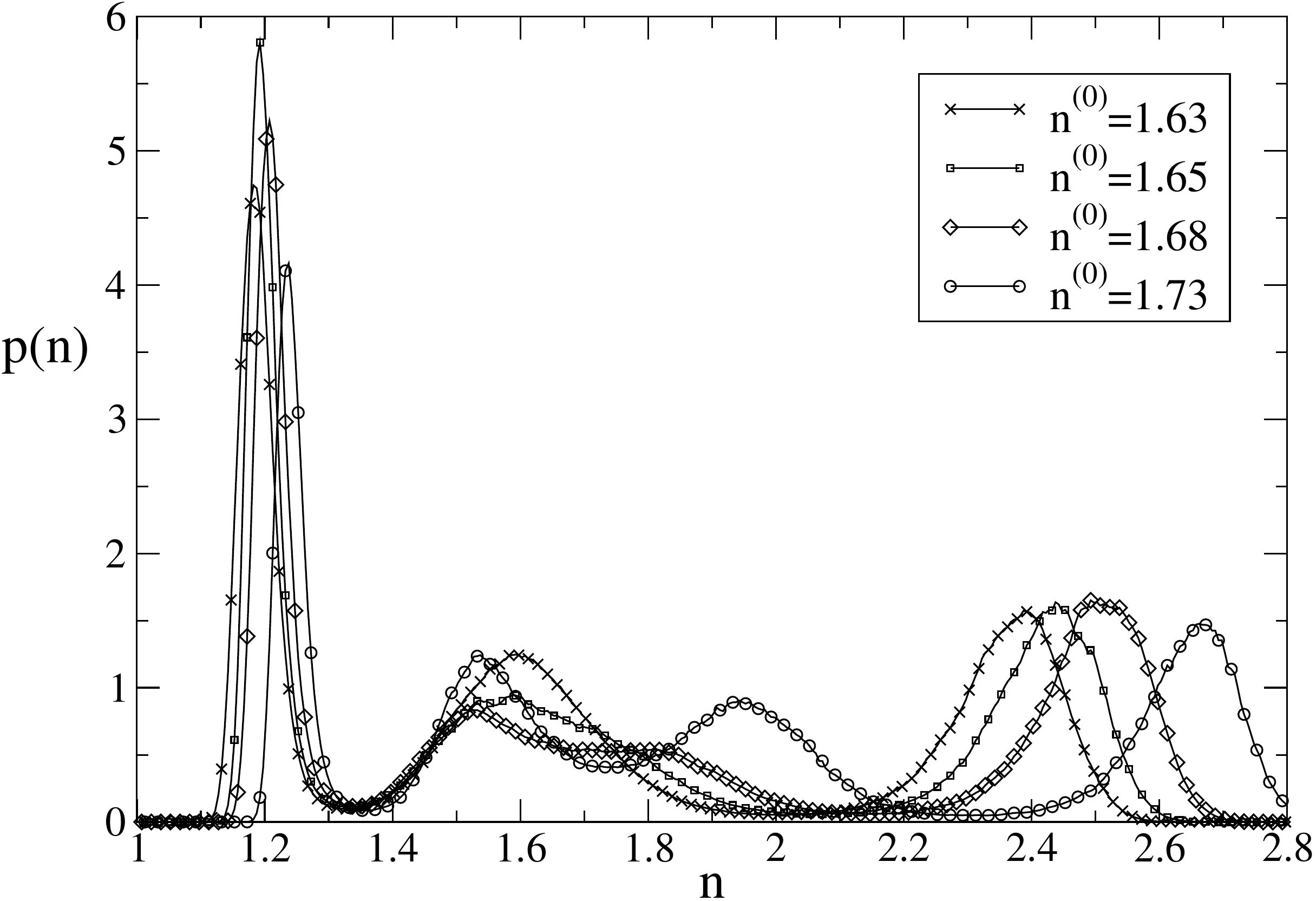}
\end{center}
\caption{Order parameter distributions through
 {\bf (a, top)} the S--SS transition and {\bf (b, bottom)} the
 SSS--SSSS transition.}
\label{fig:order_parameter}
\end{figure}

We consider first the overall phase diagram of the soft sphere system
as studied in our simulations for a system of $N=256$ particles. Fig.~\ref{fig:part_pd}a shows (empty
symbols) the boundaries of the fluid-solid (FS) coexistence region at
low densities. These boundaries are the cloud curves
coming from the low and high density regimes, respectively, and were
previously determined by us using MC phase-switch 
techniques~\cite{Wilding2010b}. 
Our focus in this paper is on the solid region at higher densities.
Here a comprehensive exploration of the
($n\p$-$\delta$) plane is impractical because of the relatively high 
computational cost of our specialized simulation technique. But we can
understand important qualitative features by following the dashed
trajectory included in Fig.~\ref{fig:part_pd}a~\footnote{In the analogous diagram of Ref.~\cite{SolWil10}, the S--SS phase
boundary was erroneously drawn slightly too low, at
$\delta=8\%$.}. Along this path, we
monitored the state of the system via the probability distribution of
the fluctuating total number density $p(n)$, which serves as an order
parameter for phase changes. Starting from the fcc solid cloud point at
$\delta=6.3\%$, we initially increased $n\p$ in a stepwise fashion
(filled circles) to $n\p=1.45$, and then switched to increasing $\delta$
at constant $n\p$ as a potentially faster route to demixing. Indeed, at
$\delta\approx 8\%$ there was a smooth change in $p(n)$ from
single to double peaked; an example of the double peaked form is shown
in Fig.~\ref{fig:order_parameter}a. The two associated phases
were identified 
as being fcc solids. As is physically reasonable, the
higher density solid (HDS) daughter  phase contains a surplus of the
smaller particles while the lower density solid (LDS) phase has more
of the larger particles; see Fig.~\ref{fig:daughters2} below.

Continuing to higher $\delta$ eventually led to spontaneous melting of
the system at $\delta=13.7\%$,  implying that the limit of metastability
with respect to a fluid-solid-solid (FSS) coexistence had been
overstepped, as is indeed predicted by our MFE calculations (see
Fig.~\ref{fig:part_pd}b).
 We therefore backtracked  slightly into the solid-solid (SS) region,
embarking on a new trajectory with increasing $n\p$ at constant
$\delta=13.5\%$. This produced a third peak in $p(n)$ at $n\p\approx
1.475$. The corresponding intermediate density solid (IDS) was again
found to be isostructural with the other two, with dominant particle
sizes between those in the HDS and LDS. Finally, increasing the overall
density to $n\p\approx 1.68$ we observed that the central IDS peak in
$p(n)$ split rather smoothly into two peaks, yielding a four peaked
structure (Fig.~\ref{fig:order_parameter}b). All four solids were again
identified as having an fcc structure.

We next compare to our theoretical MFE calculations. These used the
same parent size distribution (\ref{eq:th}) but, as explained above,
the analysis was performed for hard spheres. The reason is that 
no suitable
polydisperse model free energies are available for the soft repulsive
potential~(\ref{eq:softspheres}). Nevertheless, the qualitative
physics should be the same. Indeed, taking a comparable path
(see Appendix~\ref{app:comparable}) through
the calculated phase diagram, we find the same
features as in the simulations, as shown in Fig.~\ref{fig:part_pd}b.
Quantitatively, the fluid-solid
coexistence region is narrower, and transitions to multiple solids occur
at lower $n\p$ and $\delta$, presumably because with a hard repulsion, a
crystal can accommodate above average-sized particles less easily.

\begin{figure}[t]
\begin{center}
\includegraphics[type=pdf,ext=.pdf,read=.pdf,width=0.78\columnwidth,clip=true]{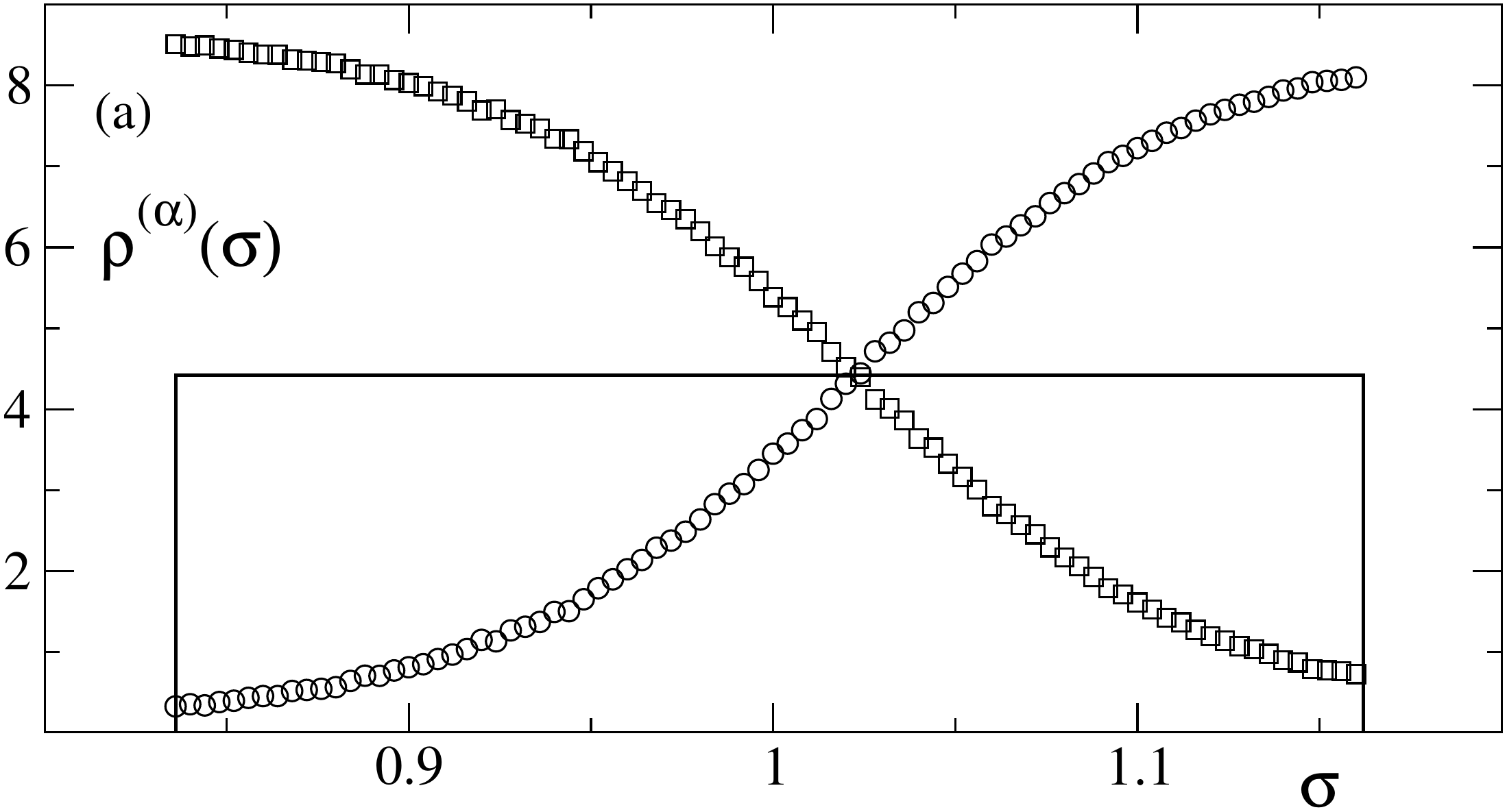}\\
\hspace*{-1.2mm}\includegraphics[type=pdf,ext=.pdf,read=.pdf,width=0.795\columnwidth,clip=true]{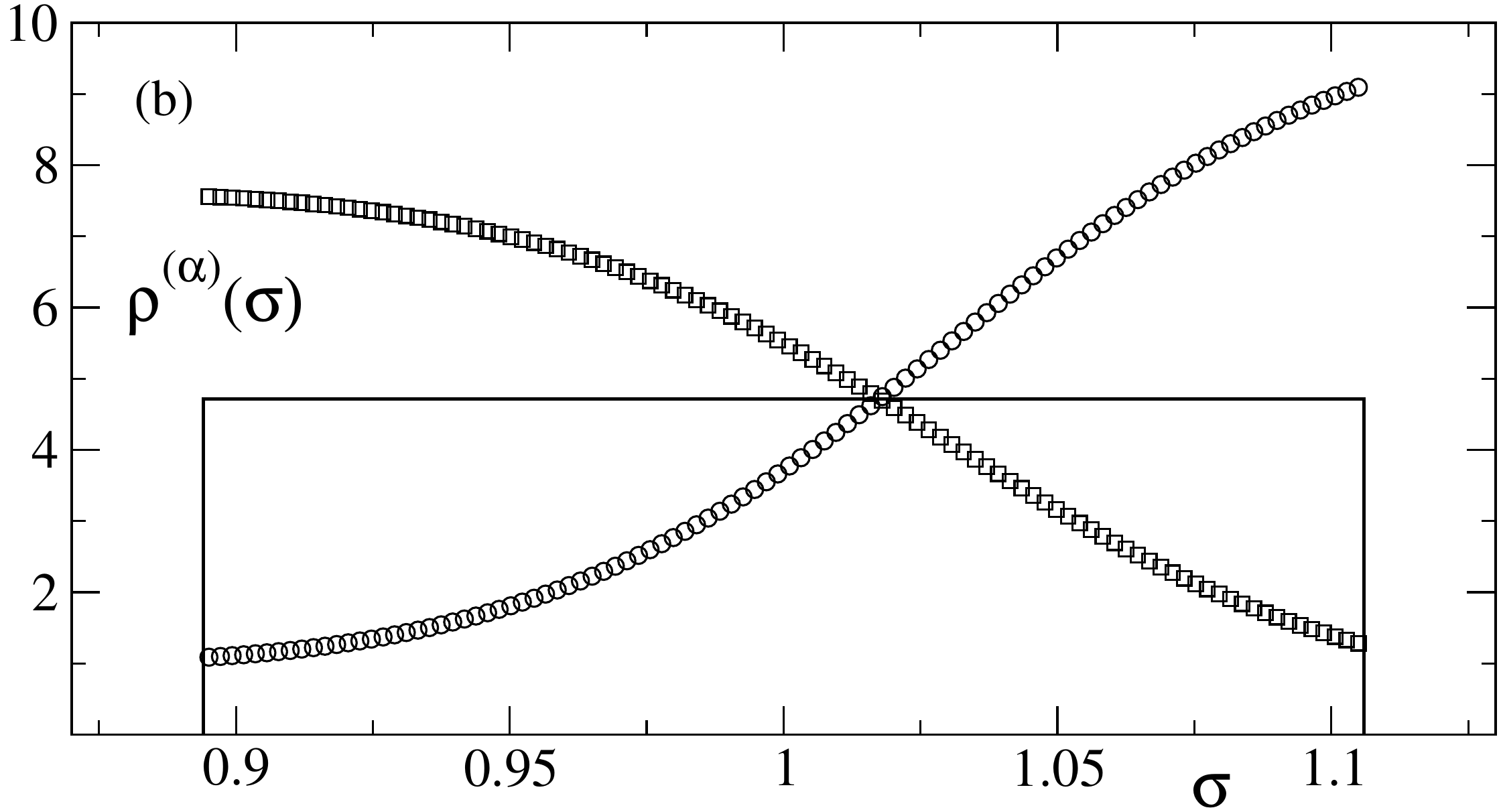}
\end{center}
\caption{Density distributions in the SS regime. {\bf (a)} Solid line: Parent density distribution at state point ($n\p=1.45,
\delta=9.5\%$), marked by the red circle in
Fig.~\ref{fig:part_pd}a. Symbols: Simulation results for the two
daughter distributions.
The associated fractional
volumes $\lambda^{(\alpha)}$ are $0.527$ (HDS, squares), $0.473$ (LDS,
circles).
{\bf (b)} MFE results at the comparable state point ($n\p=1.133$,
$\delta=6.12\%$), marked by the red circle in
Fig.~\ref{fig:part_pd}b. Fractional volumes are $0.561$ (HDS), $0.439$ (LDS).
}
\label{fig:daughters2}
\end{figure}

A key feature of the phase diagram is the absence of glassy phases.
The frustration that could otherwise engender such phases is avoided
precisely by fractionation. To illustrate this, we show in
Fig.~\ref{fig:daughters2} the density distributions for two coexisting
solids, at the state points marked by the circles in
Fig.~\ref{fig:part_pd}. The figure also shows the parent density
distribution. It is likely that if a single solid were forced to have
this size distribution at the density considered, it would indeed assume a
disordered, glassy structure. Our results show that at equilibrium,
this is avoided by effectively splitting the range of particle sizes among
two phases, allowing each phase to remain crystalline on account
of its now narrower range of particle size variation. This scenario is
then broadly in line with that proposed by Bartlett~\cite{Bartlett1998},
but the split in sizes is not ``sharp'' in the sense that
particles of a given size would be found exclusively in one phase or
the other. Such a sharp split would require
infinite differences between phases of the relevant size-dependent chemical
potentials. Apart from the general phenomenon of fractionation,
Fig.~\ref{fig:daughters2} also demonstrates good agreement between the
simulation results and the MFE predictions, with e.g.\ the crossing
point between the three density distributions in both cases located
somewhat to the right of the parental mean.

\begin{figure}[t]
\begin{center}
\includegraphics[type=pdf,ext=.pdf,read=.pdf,width=0.8\columnwidth,clip=true]{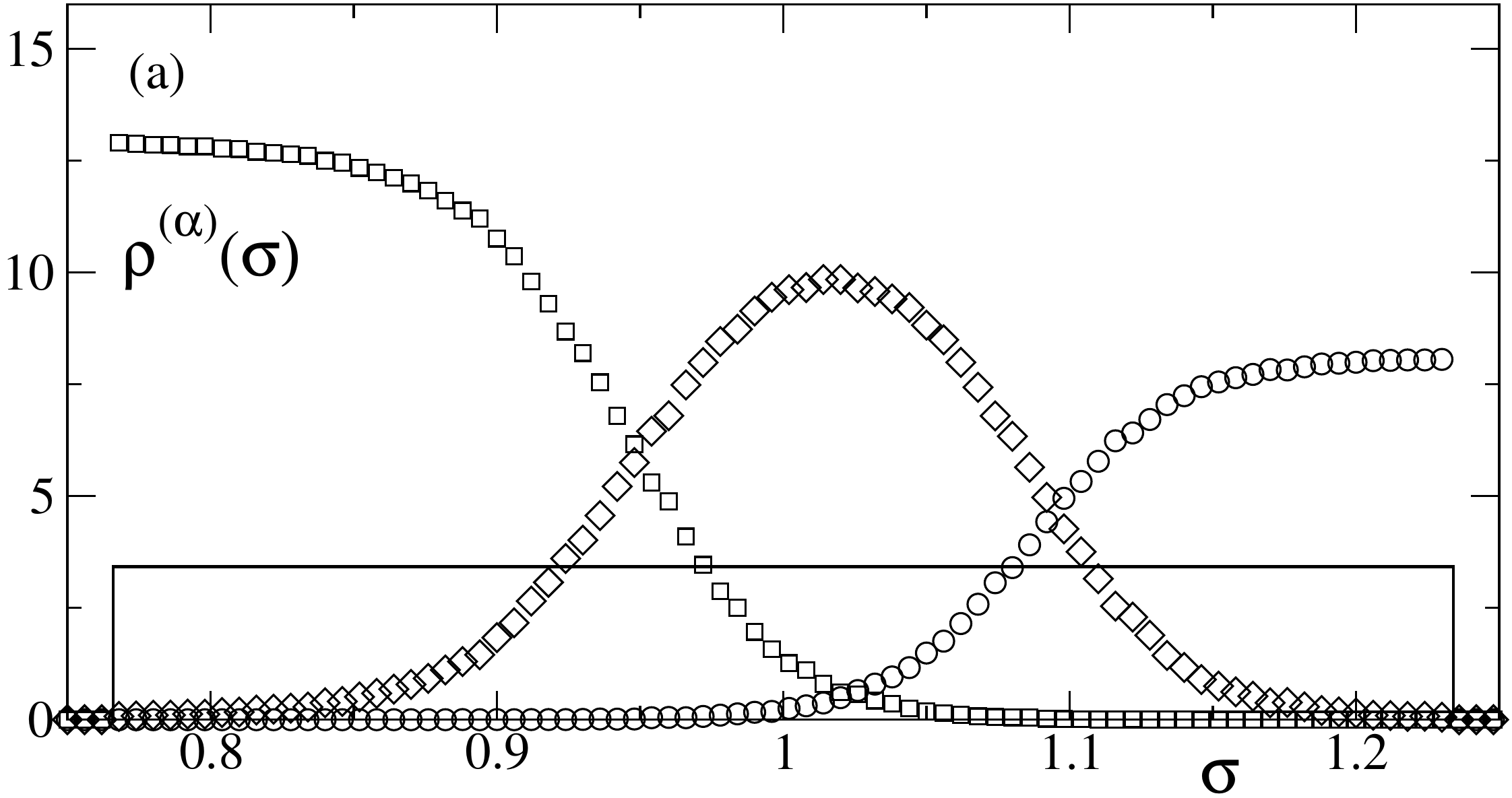}\\
\includegraphics[type=pdf,ext=.pdf,read=.pdf,width=0.8\columnwidth,clip=true]{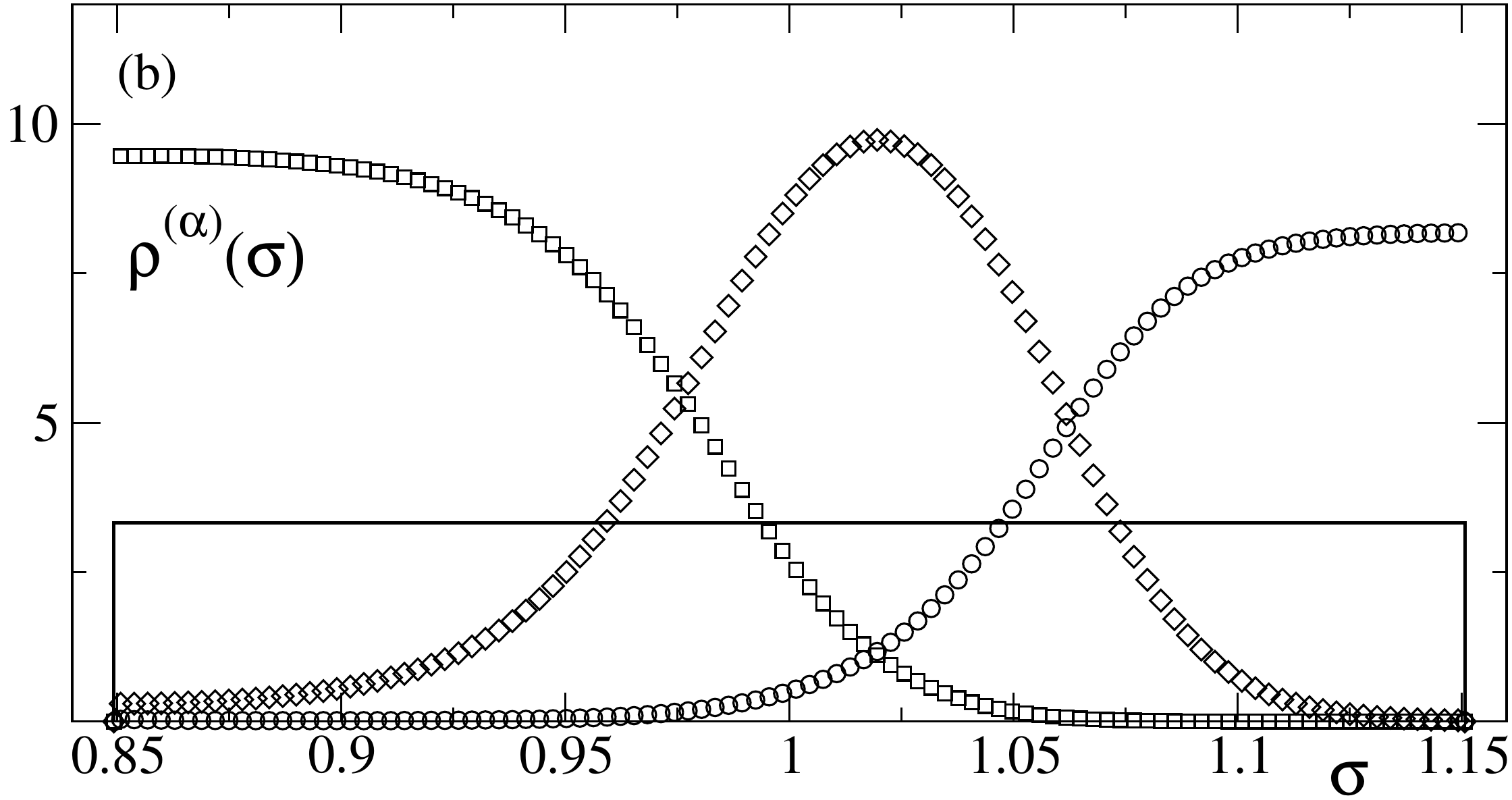}
\end{center}
\caption{
Density distributions in the SSS regime. {\bf (a)} Solid line: Parent
density distribution at state point  ($n\p=1.60,
\delta=13.5\%$), marked by the green triangle in
Fig.~\ref{fig:part_pd}a. Symbols: Simulation results for the three
daughter distributions.
 The associated fractional
volumes $\lambda^{(\alpha)}$ are  $0.267$
(HDS, squares), $0.309$ (IDS, diamonds), $0.424$ (LDS, circles).
{\bf (b)} MFE results at the comparable state point
($n\p=1.186$,
$\delta=8.7\%$), marked by the green triangle in
Fig.~\ref{fig:part_pd}b. 
Fractional volumes are $0.341$ (HDS), $0.253$ (IDS),
$0.405$ (LDS).}
\label{fig:daughters3}
\end{figure}

\begin{figure}[t]
\begin{center}
\includegraphics[type=pdf,ext=.pdf,read=.pdf,width=0.8\columnwidth,clip=true]{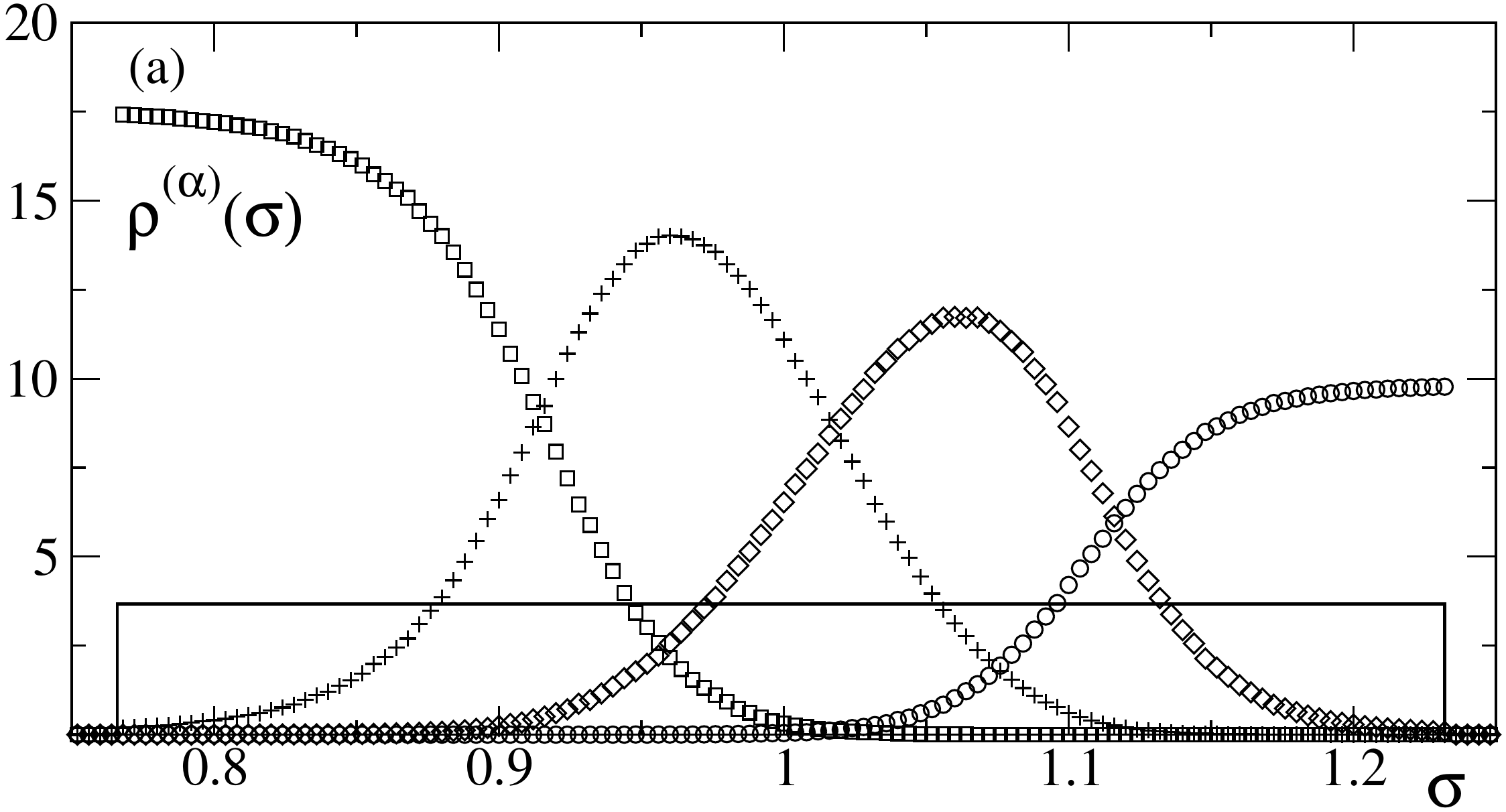}\\
\includegraphics[type=pdf,ext=.pdf,read=.pdf,width=0.8\columnwidth,clip=true]{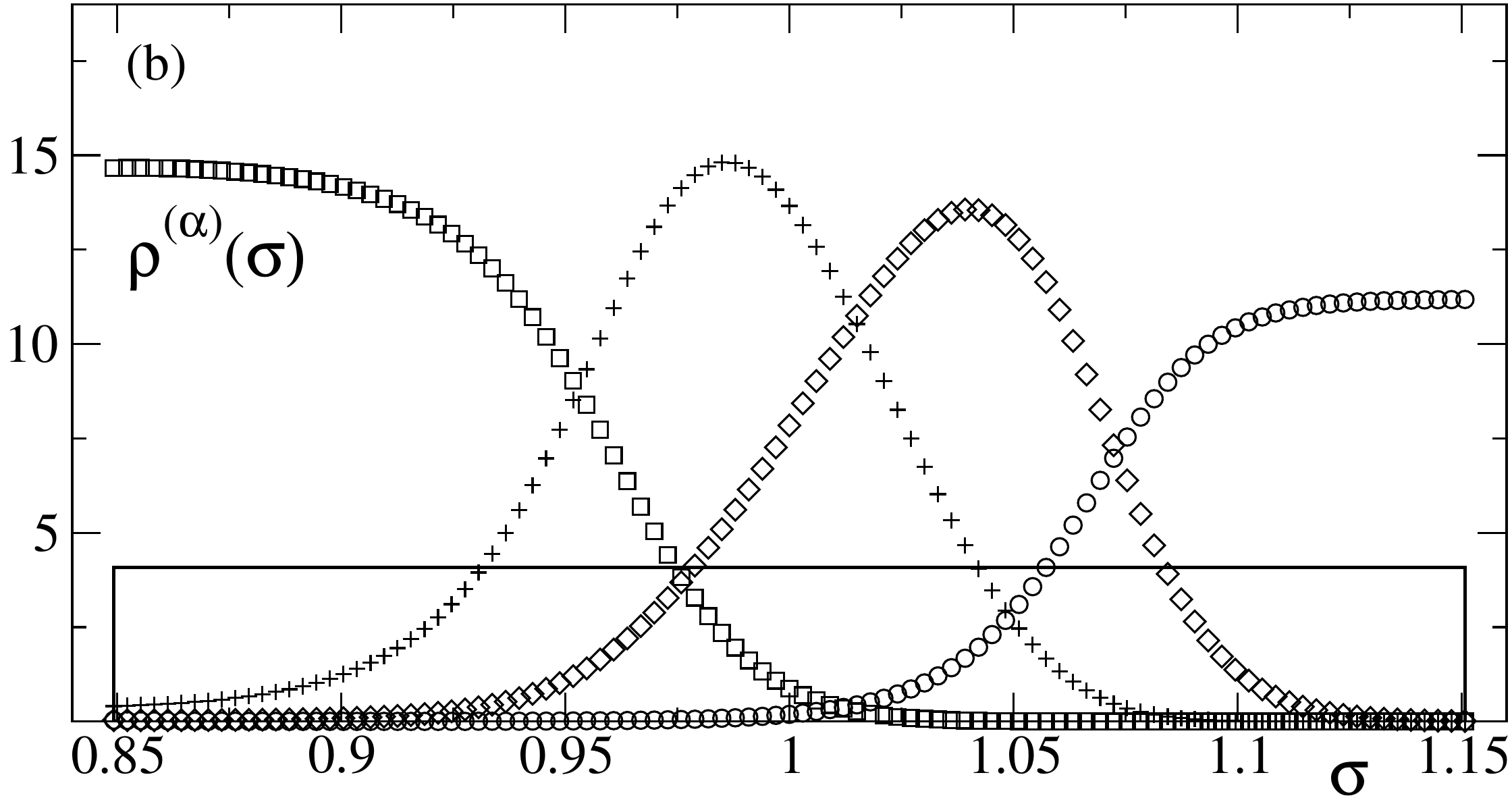}
\end{center}
\caption{
Density distributions in the SSSS regime. {\bf (a)} Solid line: Parent
density distribution at state point  ($n\p=1.73,
\delta=13.5\%$), marked by the blue square in
Fig.~\ref{fig:part_pd}a. Symbols: Simulation results for the four
daughter distributions. The associated fractional
volumes $\lambda^{(\alpha)}$ are, from left to right: 0.209 (HDS,
squares), 0.188 (IDS2, $+$), 0.232 (IDS1, diamonds), 0.373 (LDS, circles).
{\bf (b)} MFE results at the comparable state point
($n\p=1.186$,
$\delta=8.7\%$), marked by the blue square in
Fig.~\ref{fig:part_pd}b.
Fractional volumes are, from left to right, 0.273, 0.162, 0.200, 0.365.
From Ref.~\cite{SolWil10}. Copyright
American Physical Society.}
\label{fig:daughters4}
\end{figure}

As the density or parent polydispersity of a system in the SS regime
are increased further, the polydispersity in the two daughter phases
becomes unfavorably large. At this point a third solid appears that
takes up the middle of the size distribution, producing three
daughters whose size distributions are again sufficiently narrow. This
is illustrated in Fig.~\ref{fig:daughters3}. At the transition from
this SSS regime to four solids (SSSS), we then see a process
that is qualitatively similar to the S--SS transition: the middle (IDS)
phase splits into two phases, each again with a narrower size
distribution (Fig.~\ref{fig:daughters4}). It is worth emphasizing that
also in these more complicated fractionation scenarios, the agreement
between simulations for soft spheres and theory for hard spheres
remains good.

\begin{figure}[t]
\begin{center}
\includegraphics[type=pdf,ext=.pdf,read=.pdf,width=0.8\columnwidth,clip=true]{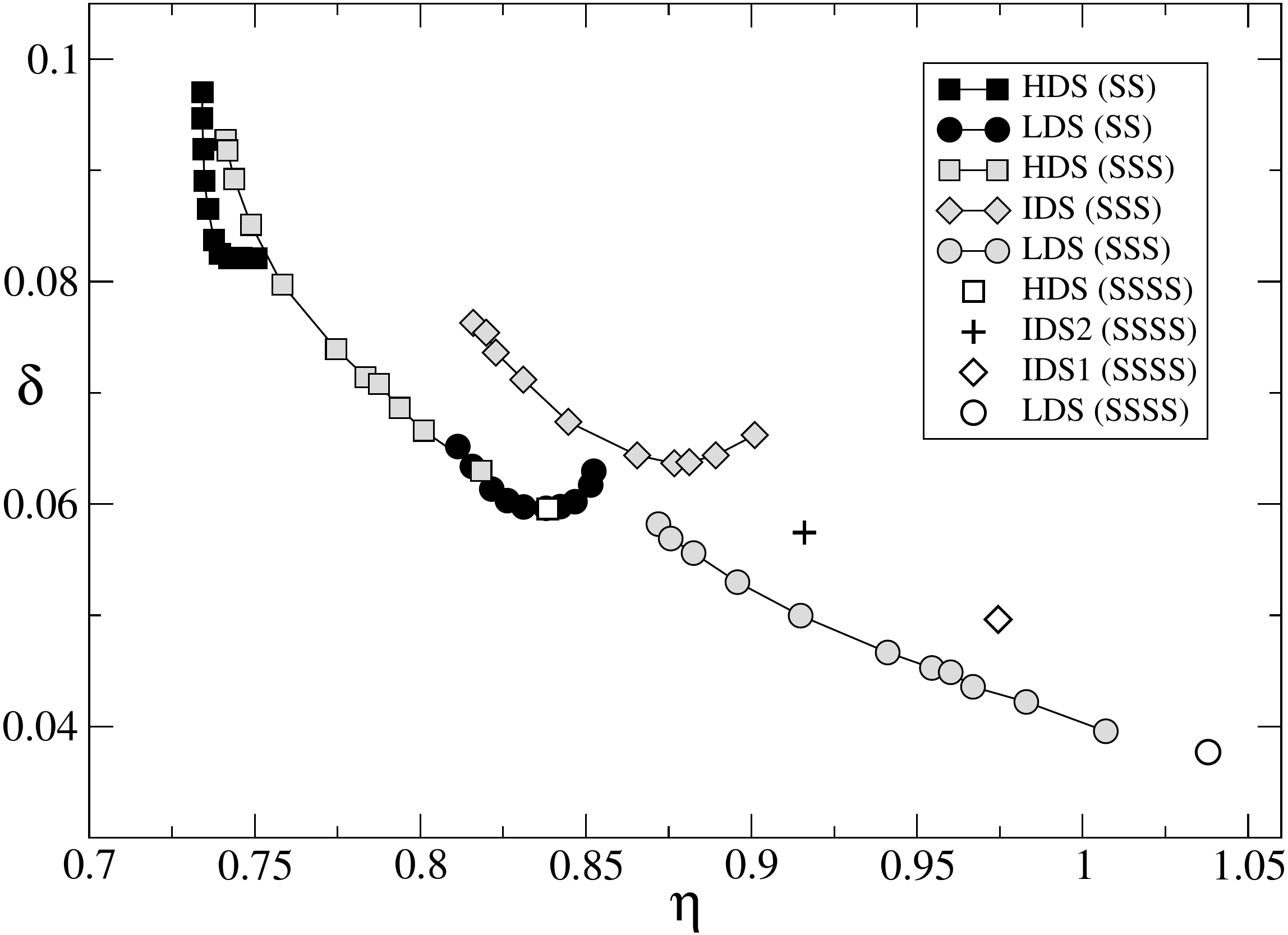}\\
\includegraphics[type=pdf,ext=.pdf,read=.pdf,width=0.8\columnwidth,clip=true]{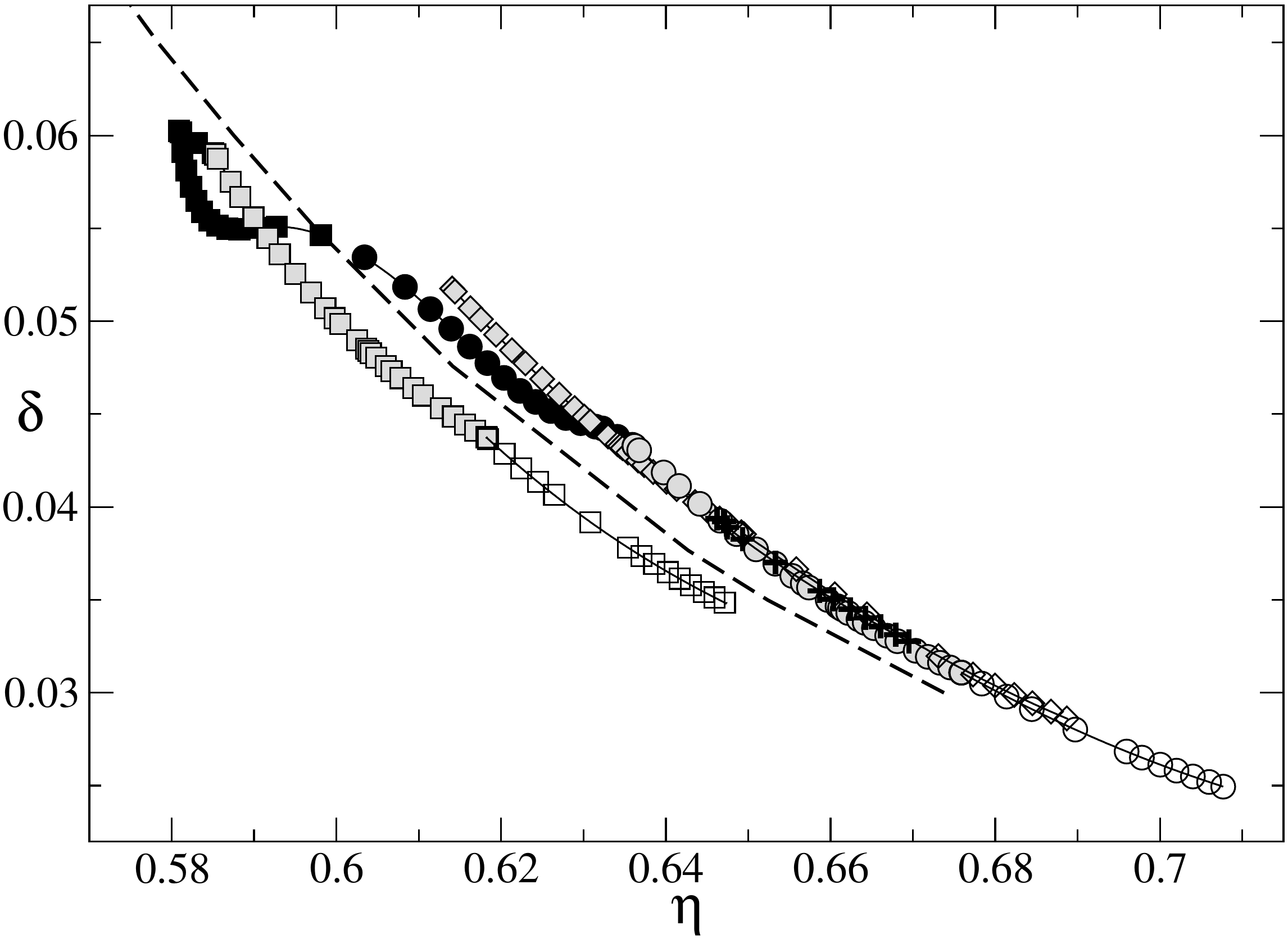}
\end{center}
\caption{{\bf (a, top)} Location of daughter phases along the vertical and
  final horizontal paths
  in the phase diagram of Fig.~\ref{fig:part_pd}a, plotted in terms of
  volume fraction $\eta$ and polydispersity $\delta$. Solid, gray and
  empty symbols refer to the SS, SSS and SSSS regions, respectively.
{\bf (b, bottom)} Analogous plot for MFE calculations for hard spheres. The
additional dashed line indicates the S--SS cloud curve for top hat size
distributions from Fig.~\ref{fig:part_pd}b.
  }
\label{fig:stability}
\end{figure}

A natural question to ask about the results so far is: what determines
the stability of solid phases, i.e.\ when do new solids appear?
Intuitively one would expect that there should be a certain threshold
in polydispersity beyond which a given single solid phase would become
thermodynamically unfavourable. This threshold should then depend on
how dense the phase is: a denser solid can accommodate less variation
in particle sizes. To test this idea quantitatively, we plot along the
path through our phase diagram the polydispersity $\delta$ versus the
volume fraction $\eta$ of all coexisting phases. The results are shown
in Fig.~\ref{fig:stability}. One sees that the coexisting phases do indeed
cluster around a line in the ($\eta,\delta$) plane, although the
clustering is clearly tighter for the MFE (hard sphere) theory. In the
plot for the latter case we also show the S--SS phase boundary from
Fig.~\ref{fig:part_pd}b as a dashed line. Recall that this is the
boundary as it applies to solids with a top hat size
distribution. Most of the coexisting phases that we find lie inside
this phase boundary, implying that with their smoother size
distributions they can tolerate a somewhat larger amount of
polydispersity. In summary, while the general picture of a line in
the volume fraction--polydispersity plane where solids become unstable holds
true, this line is broadened into a transition region by its
dependence on the shape of the size distribution. Motivated by this
finding we also experimented with other measures of polydispersity to
see whether they would reduce this dependence on distribution
shape. In particular, we considered $\delta_{2n} = [\frac{1}{2}\langle
(\sigma-\sigma')^{2n}\rangle]^{1/(2n)}/\langle\sigma\rangle$ where the
averages are over particle sizes $\sigma$ and $\sigma'$ randomly
drawn from the relevant size distribution. For
$n=1$ this gives the conventional $\delta$; for $n\to\infty$ it becomes
the difference between the largest and the smallest particle size
present, normalized by the mean size. While one may imagine the latter
quantity to be the most relevant one for determining crystal stability, we
found in practice that the clustering in the ($\eta,\delta_{2n}$)
plane becomes worse for larger $n$, with the most easily interpretable
results being the ones shown above for $n=1$.

As a final comment on Fig.~\ref{fig:stability} it is worth
highlighting that the phase with the highest density (HDS, shown by
squares) in fact always has the smallest volume fraction among the
daughter phases for a given parent. This again reflects the strong
fractionation effects: as illustrated in
Figs.~\ref{fig:daughters2}--\ref{fig:daughters4}, the HDS phase
contains the smallest particles, and this reduces its volume
fraction $\eta$ to the point where it is smaller than for all other
phases. This trend is true throughout, i.e.\ the ordering of the daughter
phases by density $n$ is always the reverse of the ordering by volume
fraction $\eta$.

\section{Criticality in transitions to multiple solids}
\label{sec:criticality}

\subsection{Order parameter distributions and fractional volumes}

In this section we discuss the nature of the transitions as our system
of polydisperse spheres fractionates into an increasing number of
solids. Our focus will be on the rather surprising finding that these
transitions can be nearly continuous in character.

Initial evidence for this claim is provided by 
Fig.~\ref{fig:order_parameter} above. This shows the distributions $p(n)$ of
the fluctuating number density in the MC simulations, with each peak
corresponding to one of the solid phases. One sees in
Fig.~\ref{fig:order_parameter}a, for the
S--SS transition, that the initial single peak splits smoothly into two
nearby peaks which then rapidly move outwards towards more clearly
separated densities. This contrasts with what one would have expected
for a first order transition, where a new peak appears at some finite
distance from the initial peak and gradually acquires more and more
weight. Such a scenario is found, along our particular path through
the phase diagram, for the SS--SSS transition (data not shown). The
SSS--SSSS transition, on the other hand, is again nearly continuous,
like the S--SS transition. This can be seen in
Fig.~\ref{fig:order_parameter}b, where the middle peak splits smoothly
into two new peaks which move apart and form the IDS1 and IDS2 phases.

\begin{figure}[t]
\begin{center}
\includegraphics[type=pdf,ext=.pdf,read=.pdf,width=0.9\columnwidth,clip=true]{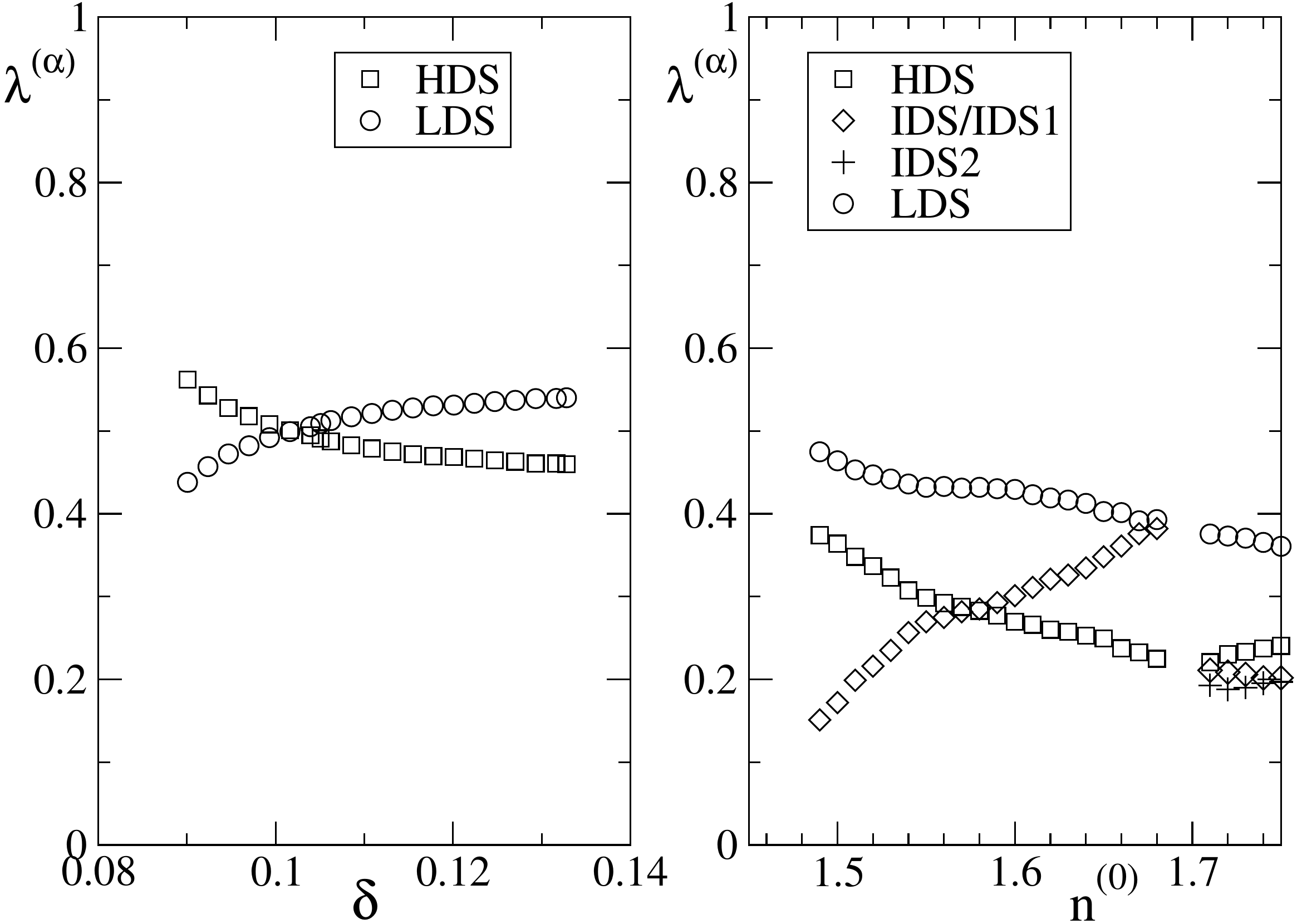}\\
\includegraphics[type=pdf,ext=.pdf,read=.pdf,width=0.9\columnwidth,clip=true]{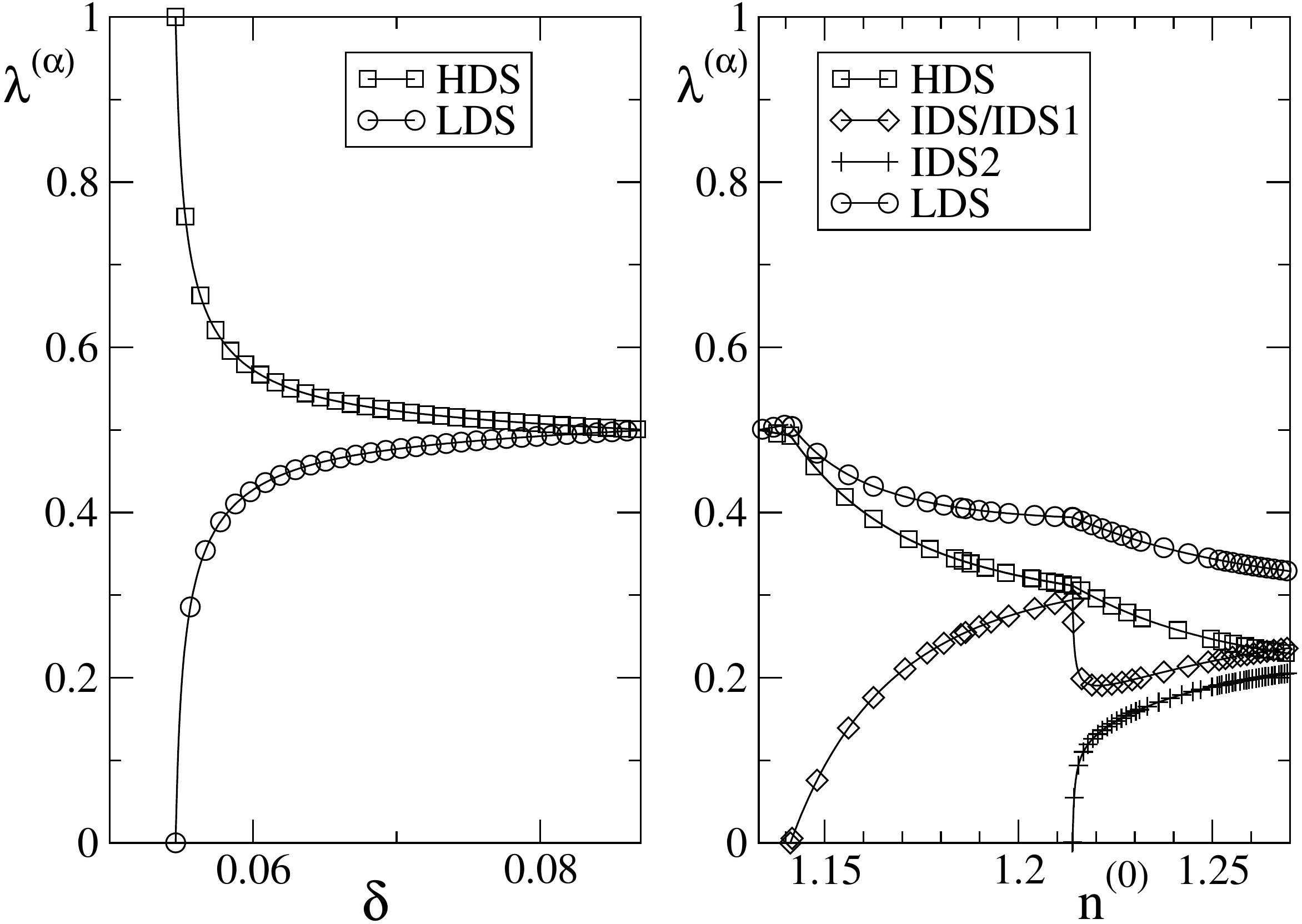}
\end{center}
\caption{{\bf (a, top)} Variations of
  $\lambda$ through transitions from single solid to multiple solids
  (SS, SSS, SSSS). {\bf (b, bottom)} Corresponding results from MFE calculations.
}
\label{fig:xis}
\end{figure}
Further evidence for nearly continuous transitions to multiple solids
is provided by the variation of the fractional phase volumes
$\lambda^{(\alpha)}$, shown in Fig.~\ref{fig:xis}. One observes that
at the S--SS transition, the fractional volume occupied by the new
phase has a strongly nonlinear variation with the parent
polydispersity. In fact, looking at the simulation results
(Fig.~\ref{fig:xis}a), where we cannot get reliable data close
to the transition, one would guess that the fractional volume of the new
phase has a discontinuous onset, as is typical of phase
transitions which are continuous in the thermodynamic sense.  The
difficulty in obtaining data close to the transition
in simulations stems from the fact that in a finite-sized system the
critical density distribution $p(n)$ has two 
peaks, so that one has to proceed some way into the two phase region before 
one can be sure that peaks observed in $p(n)$ indicate genuine phase
coexistence.
Looking at the right half of Fig.~\ref{fig:xis}, the behaviour at the
SS--SSS transition is rather different, with the fractional volume
taken up by the new phase increasing smoothly from zero in an almost
linear fashion. This is in line with expectations for a first order
transition. The SSS--SSSS transition, on the other hand, again shows nearly
continuous behaviour. As for the S--SS transition, phase coexistence
cannot be determined unambiguously from the simulation data for our
finite systems, and the data outside of the resulting gap are again
suggestive of a jump in the new fractional volume at the
transition. The MFE calculations show that there is no real jump,
rather a strongly nonlinear increase from zero, so
that the transition is close to but not fully critical.

Taken together, the above observations of the behaviour of the density
distribution $p(n)$ in the simulations, and of the variation of the
fractional phase volumes, provide strong evidence that demixing
transitions to multiple solids {\em can be} near critical. Along our
specific path through the phase diagram, it is the S--SS and SSS--SSSS
transitions that are of this type. To investigate this issue in more
detail, we now turn to characterizing the near critical properties at
the level of single phases, via appropriate correlation functions of
particle size fluctuations.

\subsection{Correlations in size fluctuations} 

To define a measure of how strongly spatially correlated size
fluctuations are in our solids, we consider first a grand canonical
setting for a single phase in a fixed volume $V$, and with imposed
chemical potentials $\mu(\sigma)$.

The fluctuating density distribution $\rho(\sigma)$ has ensemble average
$\langle\rho(\sigma)\rangle$. If we define moment densities $\rho_n = \int
d\sigma \rho(\sigma)\sigma^n$, then the normalized ensemble average size
distribution is $\langle\rho(\sigma)\rangle/\langle \rho_0\rangle$. Its
variance $\Sigma=\langle\rho_2\rangle/\langle\rho_0\rangle -
(\langle\rho_1\rangle/\langle\rho_0\rangle)^2$ sets the scale for any
particle size fluctuations. To define our correlation measure $\chi$, we
measure the mean particle size in any configuration, which is
$\rho_1/\rho_0$, and construct its variance across the ensemble. This is
then normalized by $\Sigma$ and multiplied by system volume $V$ to get a
quantity with the dimension of a volume: 
\begin{equation} \chi =\frac{V\langle\left[\Delta(\rho_1/\rho_0)\right]^2\rangle}{\Sigma}
\label{corr_vol_sim}
\end{equation} 
In the thermodynamic limit of large $V$, $\rho_0$ and $\rho_1$ have small fluctuations so one can expand
$\Delta(\rho_1/\rho_0) = (\Delta\rho_1)/\langle\rho_0\rangle -
(\Delta\rho_0)\langle\rho_1\rangle/\langle\rho_0\rangle^2$. Abbreviating
the ensemble-averaged mean size as
$\bar\sigma=\langle\rho_1\rangle/\langle\rho_0\rangle$, this gives 
\begin{equation} \chi = \frac{V\langle\left[\Delta\rho_1 - \bar\sigma
\Delta\rho_0\right]^2 \rangle}{\langle \rho_0\rangle^2 \Sigma} 
\label{xi3forMFE}
\end{equation}
or in terms of the fluctuating density distribution 
\begin{equation}
\chi = \frac{V\langle\left(\int d\sigma (\sigma-\bar\sigma)
\Delta\rho(\sigma)\right)^2 \rangle} {\langle \rho_0\rangle^2 \Sigma}
\label{corr_vol_final}
\end{equation} 
 The denominator here could also be written as $\langle\rho_2\rangle \langle\rho_0\rangle -
\langle\rho_1\rangle^2$.

To motivate further the above definition of our measure of
correlations $\chi$, one can express it via correlation functions of
the full spatially-resolved density $\rho(\rv,\sigma)$. The
fluctuations of the latter can be expressed in terms of the pair correlation
function $g_{\sigma\sigma'}(\rv)$ between particles of sizes $\sigma$
and $\sigma'$ as~\cite{HanMcD86}
\begin{eqnarray} \langle
\Delta\rho(\rv,\sigma)\Delta\rho(\rv',\sigma')\rangle &=&
\langle\rho(\sigma)\rangle \delta(\rv'-\rv)\delta(\sigma'-\sigma) \nonumber
\\ &&{}+\langle\rho(\sigma)\rangle \langle\rho(\sigma')\rangle
[g_{\sigma\sigma'}(\rv'-\rv)-1] \nonumber \end{eqnarray} 
 So the numerator of (\ref{corr_vol_final}) is, using $\Delta\rho(\sigma) = V^{-1}\int d\rv
\Delta\rho(\rv,\sigma)$, 
 \begin{eqnarray} V^{-1}\int
\!d\rv\, d\rv'd\sigma\,d\sigma' (\sigma\!-\!\bar\sigma)
(\sigma'\!-\!\bar\sigma)
\langle\Delta\rho(\rv,\sigma)\Delta\rho(\rv',\sigma')\rangle = \nonumber\\
\langle \rho_0 \rangle \Sigma \!+\! \int \!d\rv\,d\sigma\,d\sigma'
(\sigma\!\!-\!\!\bar\sigma) (\sigma'\!\!-\!\!\bar\sigma)
\langle\rho(\sigma)\rangle
\langle\rho(\sigma')\rangle [g_{\sigma\sigma'}(\rv)\!-\!1] \nonumber
\end{eqnarray}
This shows that our definition of $\chi$ is physically reasonable: it
is the volume integral of a correlation function that measures
the spatial correlations of fluctuations in particle size away from the
ensemble mean. We will therefore also refer to $\chi$ as the size
fluctuation susceptibility. Note that the trivial first term above makes a
contribution of $1/\langle \rho_0\rangle$ to $\chi$. This is the unit
volume per particle and of order
unity in the density range we are considering. We will see below
that it is negligible compared to the main contribution from the
correlation function integral.

Some care is needed when relating the susceptibility $\chi$ as defined
above to a length scale $\xi$ for the spatial correlations of size
fluctuations.  Away from criticality, and in $d$ spatial dimensions,
then since the correlation function being integrated decays on a
spatial scale of $\xi$, one estimates $\chi\sim\xi^d$. This is the
identification we made previously~\cite{SolWil10}. At criticality, on
the other hand, the correlation function appearing above will have a
spatial power law decay with $|\rv|^{-d+2-\eta}$ up to the cutoff, and
hence the susceptibility scales as $\chi\sim\xi^{2-\eta}$ where $\eta$
is the standard critical exponent (and not, as elsewhere in the paper,
the volume fraction).

One can show that, for large systems, the size fluctuations we are
considering are the same in all reasonable ensembles, for example a
semi-grand canonical ensemble where particle number $N$ is fixed and
the volume $V$ can fluctuate. In this case the factor $V$
in (\ref{corr_vol_sim}) is replaced by $N/\langle \rho_0\rangle$ to give
\begin{equation} \chi =
\frac{N\langle\left[\Delta(\rho_1/\rho_0)\right]^2\rangle}
{\langle\rho_2\rangle - \langle\rho_1\rangle^2/\langle\rho_0\rangle}
\end{equation} 
and this is the method we use to extract $\chi$ from simulation data.
For the theoretical calculations, we employ (\ref{xi3forMFE}) and
extract the (co-)variances of the fluctuations of the moment densities
$\rho_0$, $\rho_1$ from the appropriate curvature matrix of the moment
free energy~\cite{Sollich2001}.

Note finally that in the context of experiments on colloids a
canonical ensemble, with fixed particle number $N$, volume $V$ and
parent size distribution, would be the most natural
description. For a single phase, the mean size is then fixed and no
size fluctuations occur. But $\chi$ can still be defined in terms
of the pair correlation function $g_{\sigma\sigma'}(\rv)$ as described
above, provided the spatial integration over $\rv$ is cut off at some
distance much larger than the correlation length but much smaller than the
system size. This eliminates the contribution from the nonzero
values $g_{\sigma\sigma'}(\rv)-1 = O(1/N)$ that remain at larger $\rv$
when the total particle number $N$ is fixed~\cite{HanMcD86}.
Once several phases appear, each phase has fluctuating particle
numbers and volume, but one can check that the size fluctuations in
each phase, and hence the size fluctuation susceptibility, are as would be
calculated for single phases in the grand canonical ensemble.

\begin{figure}[h]
\begin{center}
\includegraphics[type=pdf,ext=.pdf,read=.pdf,width=0.82\columnwidth,clip=true]{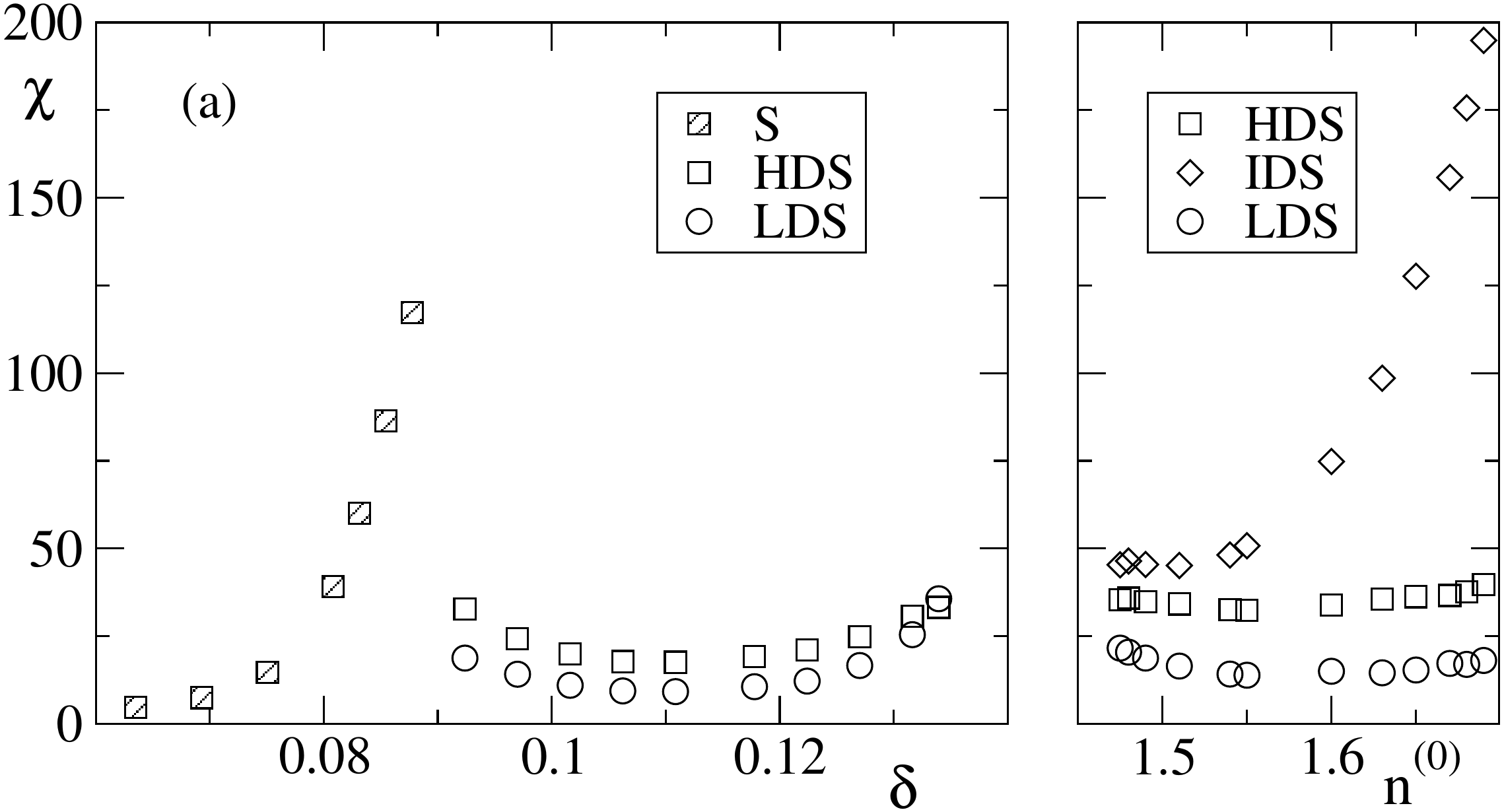}\\
\includegraphics[type=pdf,ext=.pdf,read=.pdf,width=0.82\columnwidth,clip=true]{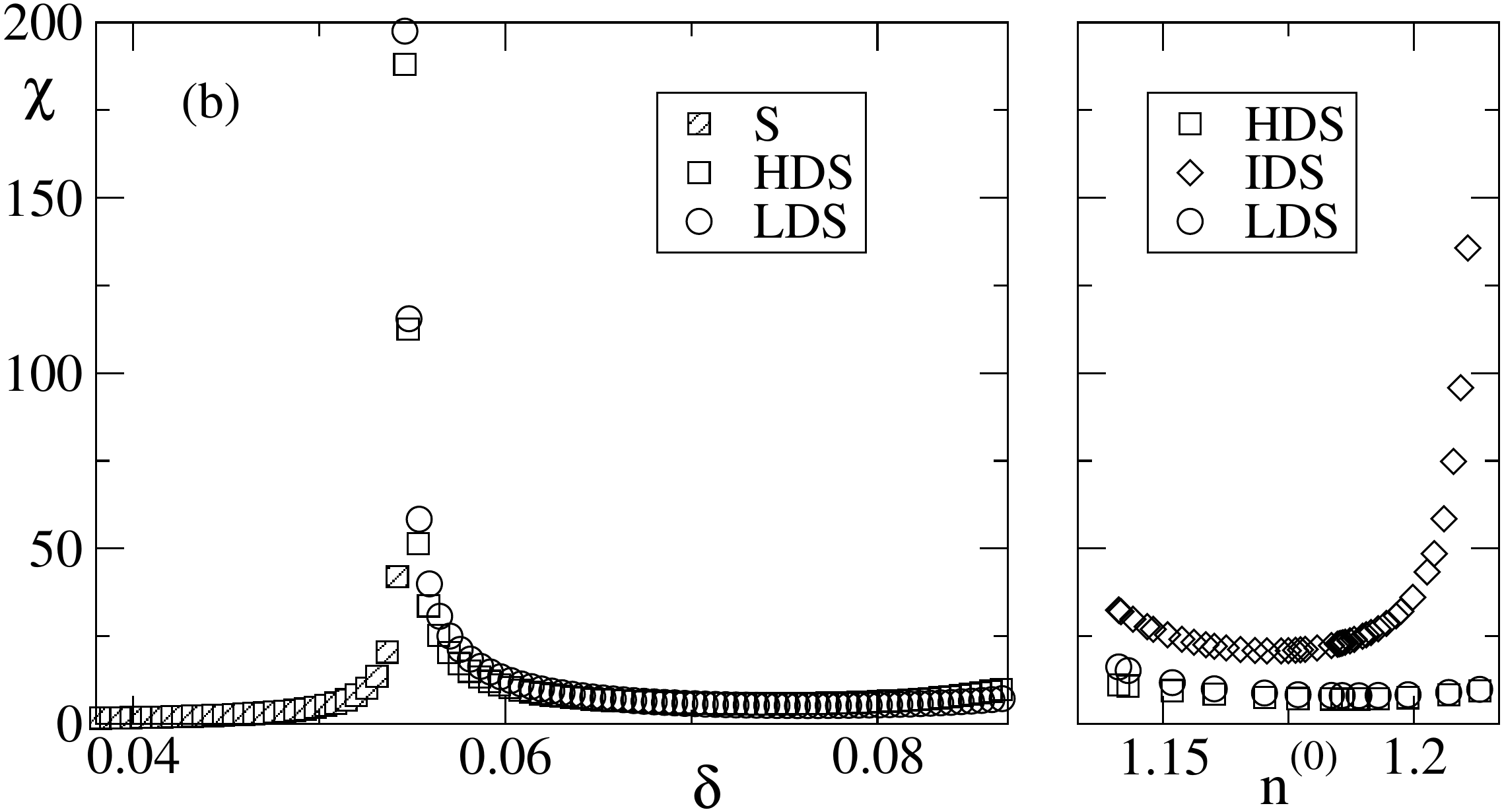}
\end{center}
  \caption{Size fluctuation susceptibility $\chi$ in the solid phases encountered along
  the phase diagram trajectories of Fig.~\ref{fig:part_pd}. {\bf (a)}
  Simulations, {\bf (b)} MFE calculations. From Ref.~\cite{SolWil10}. Copyright
American Physical Society. 
}
\label{fig:fluctuation}
\end{figure}

Having defined how we will quantify the strength of correlations in
spatial particle size fluctuations, we show in
Fig.~\ref{fig:fluctuation} results for $\chi$ along the vertical and
final horizontal paths through the phase diagrams of
Fig.~\ref{fig:part_pd}. One observes that $\chi$ grows large near the
transitions to two and four solids, confirming their near continuous
character. In the latter case, the splitting of the middle peak seen
earlier in $p(n)$ suggests that the new solids arise out of the IDS
phase, and this is consistent with large fluctuations occurring (see
Fig.~\ref{fig:fluctuation}) only in this phase and not the HDS or LDS.
The MFE predictions are, again, in good qualitative accord with the
simulation data.
%

To summarize our observations in this section, the behaviour of
fractional phase volumes and of the order parameter distributions
$p(n)$ suggested that phase transitions to multiple solid phases can
be near critical in nature; for our path through the phase diagram
this applies to the S--SS and SSS--SSSS transitions. We proposed the size
fluctuation susceptibility $\chi$ as a quantitative measure of the
range of correlations in the spatial fluctuations of particle
sizes. Results for this from both simulations and MFE calculations
then demonstrated that these transitions are indeed close
to critical, being characterized by values of $\chi$ far above the
unit volume per particle.

That such critical or near critical transitions from one to several
solids might occur is plausible given that S--SS critical points are
observed also in simulations of binary hard sphere
mixtures~\cite{KraFre91}. The free energy expression for polydisperse
hard spheres that we use in our MFE calculations was devised by
Bartlett~\cite{Bartlett97} on the basis of free energies fitted to
these binary mixture simulations. The polydisperse system must then
``inherit'' the existence of critical points, though not in any
trivial way. For example, the polydisperse system has many more
degrees of freedom for fluctuations in its size distribution, and one
can show from this that spinodal densities are always lower in the
polydisperse than in the corresponding binary case.

It is worth stressing that even in transitions involving multiple
solid phases (SSS--SSSS), criticality is essentially a single-phase property.
Indeed, we have found above quite distinct values of $\chi$ in the
three coexisting solid phases before the transition to four solids. In
the simulations, we have only a single phase in the simulation box for
most of the time, which emphasizes further that $\chi$ is determined
from the properties of this single phase. To be more precise, there
{\em is} an effect of the presence of other phases: the lever rule
forces the density distributions of all phases to add up to the
parent, and this provides constraints on the chemical potentials
$\mu(\sigma)$. Once we know these chemical potentials, however, we can
determine $\chi$ individually for every phase, independently of the others. 

We next ask what features of a given size distribution make it undergo
a critical or near critical transition to multiple solids. Having
recognized that criticality is a single-phase property, we focus in
this enterprise on
the cloud point for the transition S--SS from a single solid to two
fractionated solids.

\subsection{Predicting criticality}

The question of determining whether the S--SS transition from a parent
with a given size distribution is close to critical can be cast in
quantitative terms as follows: how large is the size fluctuation
susceptibility $\chi$ at the S--SS cloud point? We investigate this
using MFE calculations for the hard sphere case; precise simulation studies
would inevitably require finite-size scaling to larger system sizes
than we can access using our computational resources.

As explained in Sec.~\ref{sec:MFE}, the free energy expression that we
use for polydisperse hard spheres have excess contributions that
depend only on the moment densities $\rho_i$ with $i=0,1,2,3$, defined
by the weight functions $w_i(\sigma)=\sigma^i$. One can then show in
generality (see e.g.~\cite{SolWarCat01}) that the criterion for a
spinodal, where a phase becomes unstable to local density
fluctuations, involves these moments as well as those defined by the
second-order weight functions $w_i(\sigma)w_j(\sigma)=\sigma^{i+j}$,
giving in our case moments up to $\rho_6$. For a given particle size
distribution, all the ratios $\rho_1/\rho_0, \ldots, \rho_6/\rho_0$
are fixed and the density $\rho_0\equiv n$ at the spinodal can be found from
the spinodal criterion.  The additional condition for a spinodal point
also to be a critical point involves in addition the third-order
weight functions $w_i(\sigma)w_j(\sigma)w_k(\sigma)=\sigma^{i+j+k}$,
which produce moments up to $\rho_9$. Inserting the spinodal density,
the exact critical point condition resulting from our model free
energies is then some function of $\rho_1/\rho_0, \ldots,
\rho_9/\rho_0$. These are the 1st to 9th moments of the normalized
size distribution, and so whether a parent phase with a given size
distribution will exhibit a critical S--SS transition or not depends
only on these moments.

Unfortunately, because the solid free energies we use are derived from
fits to simulation data~\cite{Bartlett97}, the critical point
condition that results is far too complicated to allow for any
analytical progress. We therefore proceed initially by solving the
condition numerically for a range of parent size distributions of
interest. The first case to consider is evidently the top hat
distribution studied throughout the paper so far. Here there is only a
single parameter to vary, namely the polydispersity $\delta$. We solve
for each $\delta$ the spinodal condition to find the spinodal density,
and then evaluate the critical point condition at this density. It
turns out that there is indeed a critical point in the phase diagram,
at ($n\p=1.1669,\delta=0.0472)$. It is marked in
Fig.~\ref{fig:part_pd}b, and lies close to the path through the
phase diagram that we have considered above. This rationalizes why the
S--SS transition along this path is near critical, with a large value
of $\chi$: at the critical point itself, we would have found
$\chi$ diverging to infinity at the transition.

The situation with regard to the shape of the parent distribution is
not trivial, however. For example, in previous work we considered both
triangular and Schulz distributions~\cite{FasSol04}, and found no
critical points on the S--SS cloud curve in the physically relevant
ranges of density and polydispersity. To get more insight, we consider
next families of parent distributions where we can tune both the
width, as measured by $\delta$, and the shape. Generalizing from the
top hat case studied above, we look first at ``slanted top hat''
parents where the size
distribution is $f(\sigma) = A+B\sigma$ in some interval $\sigma_-\leq
\sigma\leq \sigma_+$, and zero otherwise. We adjust $A$, $B$,
$\sigma_-$ and $\sigma_+$ so that $f(\sigma)$ is normalized, has mean
1 as before, and the desired value of $\delta$. This leaves one degree
of freedom, which we express via the slant ratio
$R=f(\sigma_+)/f(\sigma_-)$, with $R=1$ giving back the simple top hat
distribution.

\begin{figure}[h]
\begin{center}
\includegraphics[type=pdf,ext=.pdf,read=.pdf,width=0.82\columnwidth,clip=true]{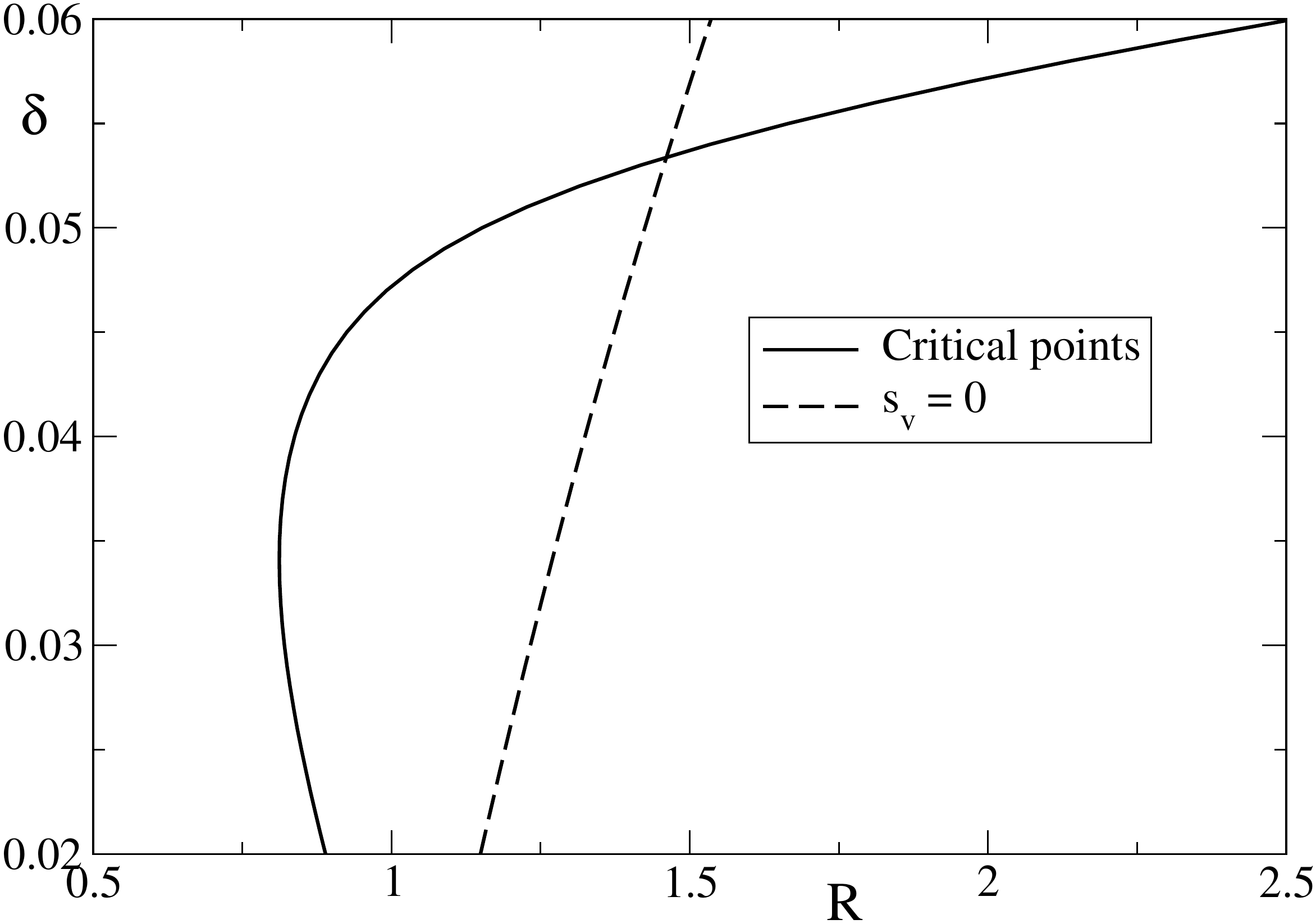}
\end{center}
  \caption{Critical polydispersity $\delta$ versus slant ratio $R$ for
    slanted parents. Dashed: approximation from $s_v=0$.}
\label{fig:critical_slanted}
\end{figure}

Proceeding as for the top hat parent, we can now determine numerically
for fixed slant ratio $R$ the critical value (if any) of $\delta$, or
vice versa. In the resulting Fig.~\ref{fig:critical_slanted} we
observe that whether or not there are critical points for a given
parent shape depends on $R$: for $R$ below around 0.81, no critical
points appear; for slightly larger values, two critical points can
exist in the phase diagram, and for values of $R$ around unity and
above we generically find one critical point.

\begin{figure}[h]
\begin{center}
\includegraphics[type=pdf,ext=.pdf,read=.pdf,width=0.82\columnwidth,clip=true]{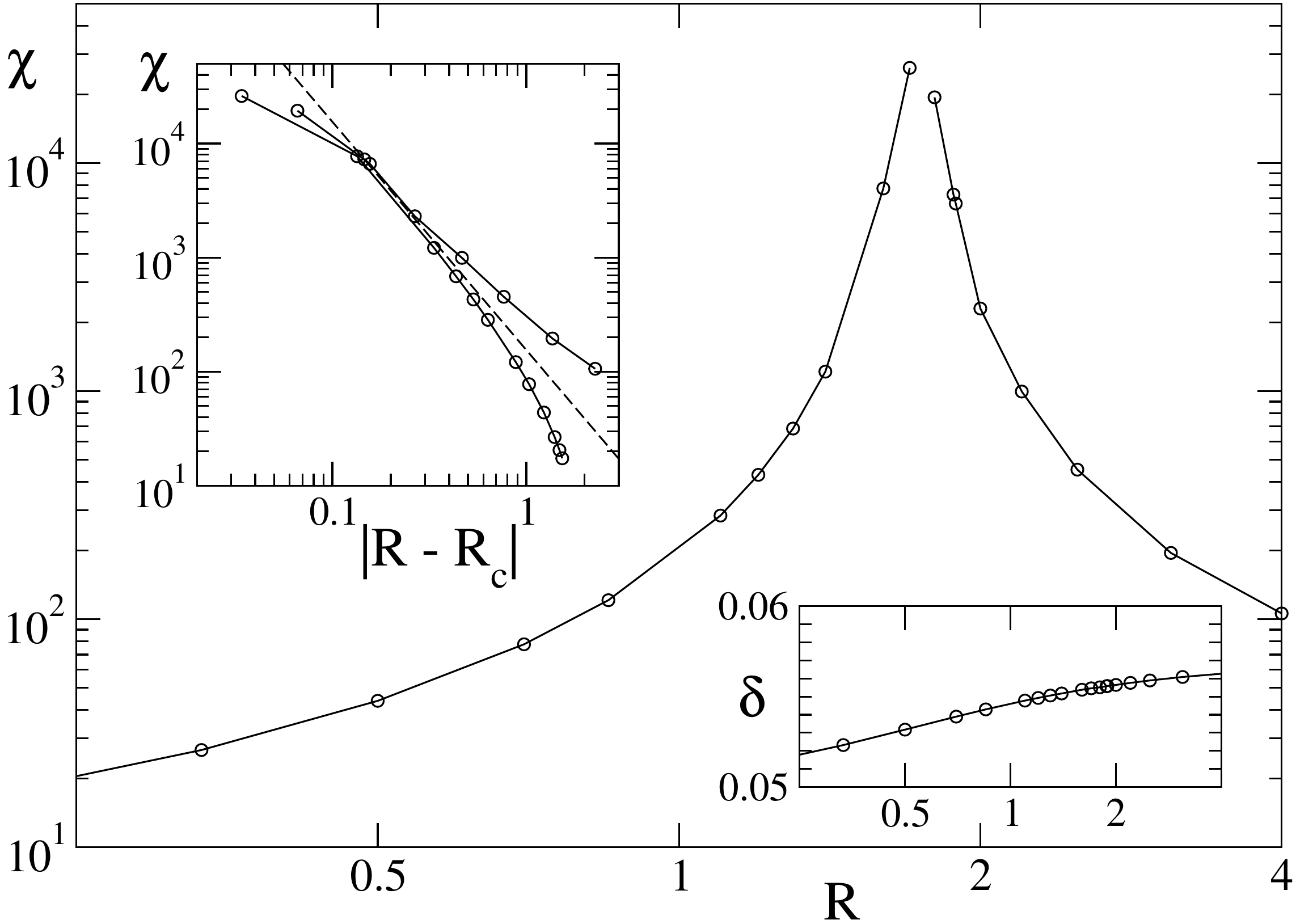}
\end{center}
  \caption{Size fluctuation susceptibility $\chi$ at S--SS cloud point
    for slanted parents with density $n\p=1.133$, versus slant ratio
    $R$. Bottom inset: $\delta$
    at the cloud point. Top inset: Semi-log plot of $\chi$ vs the deviation
    from the critical slant ratio, $|R-R_{\mathrm c}|$. The dashed
    line is a power law with exponent $-2$.}
\label{fig:chi_vs_R_slanted}
\end{figure}

So far we have asked what marks out parent size distributions that
have critical S--SS transitions, which corresponds to $\chi=\infty$ at
the cloud point. Here we digress slightly to ask how $\chi$ at the
cloud point then varies as we move away from the critical parent
shape. Data from MFE calculations are shown for this in 
Fig.~\ref{fig:chi_vs_R_slanted}, where we consider parents with fixed
density $n\p=1.133$ as on the vertical path in
Fig.~\ref{fig:part_pd}b. For given slant ratio $R$ we find the
polydispersity $\delta$ at the cloud point; see the bottom inset of
Fig.~\ref{fig:chi_vs_R_slanted}. The main plot displays the resulting cloud
point value of the susceptibility $\chi$ against $R$. It is seen to
diverge as a critical value $R=R_{\mathrm c}$ is approached, and
indeed by solving the critical point criterion for the given parent
density we find a single such critical value, $R_{\mathrm c} = 1.734$.
%
%
This means that if we had considered a parent with this slanted shape, 
we would have seen -- within our MFE calculations for hard spheres -- 
a fully critical S--SS transition on the vertical path in
Fig.~\ref{fig:part_pd}b.

The top inset of Fig.~\ref{fig:chi_vs_R_slanted} plots the
susceptibility $\chi$ versus the distance from the critical parent
shape. The data are consistent with a divergence as $\chi\sim
|R-R_{\mathrm c}|^{-2}$, except for the points nearest $R_{\mathrm c}$
either side, where our numerics become unreliable. That exponent value
may seem surprising at first: for our mean field free energy, the
susceptibility for models in the Ising universality diverges as $\chi
\sim |T-T_{\mathrm c}|^{-\gamma}$ with $\gamma=1$. But $R$ smoothly
changes the parent shape, and the latter is analogous to the Ising
magnetization $m$. We should then identify $|R-R_{\mathrm c}|$ with $m$ and
this leads to the scaling $\chi \sim m^{-\gamma/\beta} \sim
|R-R_{\mathrm c}|^{-\gamma/\beta}$. For our mean field free energy
this gives an exponent value $\gamma/\beta=1/(1/2)=2$, exactly as
observed. A more accurate theory which captures the non-mean field
Ising singularities would then be expected to give in $d=3$ the
exponent $\gamma/\beta=\delta-1\approx 3.9$.

The bottom inset of
Fig.~\ref{fig:chi_vs_R_slanted} shows the value of 
the polydispersity $\delta$ at the cloud point against the slant ratio
$R$, for the same fixed parent density as in the main plot. The
variation in $\delta$ is very small, between around 0.052 and 0.056, even
though the parent shape changes quite dramatically from $R=1/4$ to
$R=4$. This is in line with the expectation that $\delta$ is the main aspect
of the size distribution that determines solid stability.

\begin{figure}[h]
\begin{center}
\includegraphics[type=pdf,ext=.pdf,read=.pdf,width=0.82\columnwidth,clip=true]{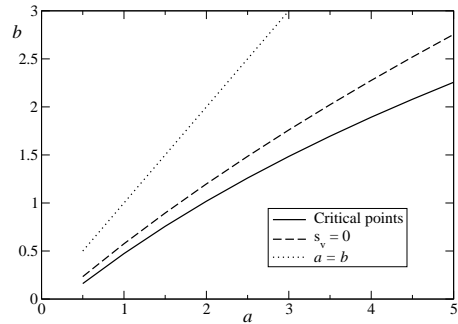}
\end{center}
  \caption{Critical line in the $(a,b)$-plane for Beta 
size distributions. The density is fixed to $n=1.133$ and the
polydispersity $\delta$ is determined from the spinodal condition; it
varies very little over the range shown, from 0.0559 to 0.0576.
The line resulting from the criterion $s_v=0$ is also shown.
Both lie significantly below the line for symmetric distributions ($a=b$).
}
\label{fig:beta_a_b_plane}
\end{figure}

Returning now to the question of what determines whether a given
particle size distribution will produce a critical S--SS transition,
we broaden our investigation to a wider class of distributions, namely
the Beta distributions. These are of the form $f(\sigma)\propto
(\sigma-\sigma_-)^a (\sigma_+-\sigma)^b$ in some interval $\sigma_-\leq
\sigma\leq \sigma_+$, and zero otherwise. The values of the smallest
and largest 
sizes $\sigma_-$ and $\sigma_+$ and the proportionality coefficient
are again adjusted to make $f(\sigma)$ normalized with unit mean and
standard deviation $\delta$. The advantage of Beta distributions is
that with their two shape parameters, $a$ and $b$, they are more
flexible than e.g.\ the slanted top hat parents from above. In
particular, they can interpolate from distributions with fairly sharp
cutoffs at the extreme sizes -- for low $a$ and $b$, where in
particular $a=b=0$ gives back a top hat distribution -- to ones with
almost Gaussian shape (large $a$ and $b$) where the cutoffs are in the
far tails of the distribution.

\begin{figure}[h]
\begin{center}
\includegraphics[type=pdf,ext=.pdf,read=.pdf,width=0.82\columnwidth,clip=true]{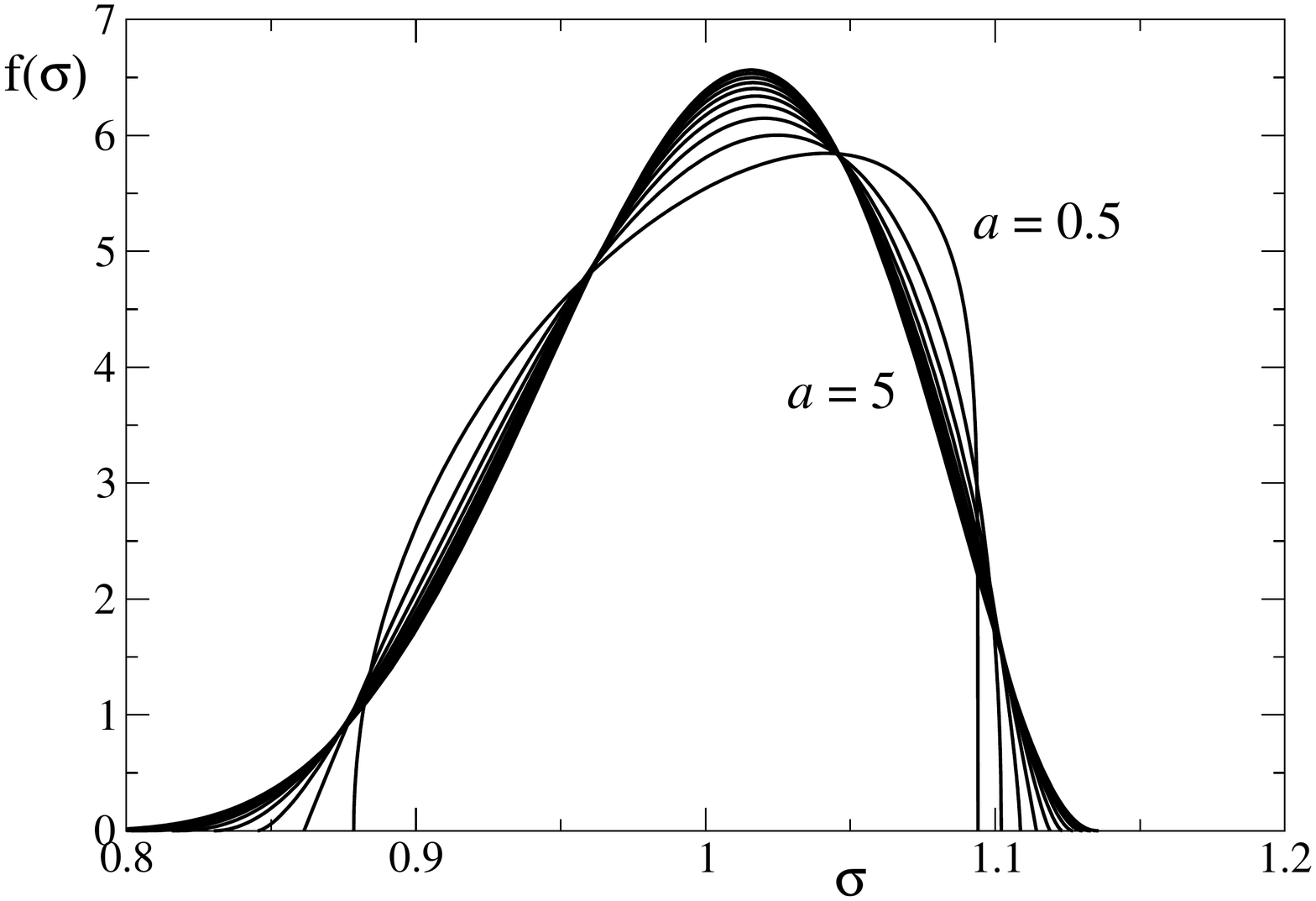}
\includegraphics[type=pdf,ext=.pdf,read=.pdf,width=0.795\columnwidth,clip=true]{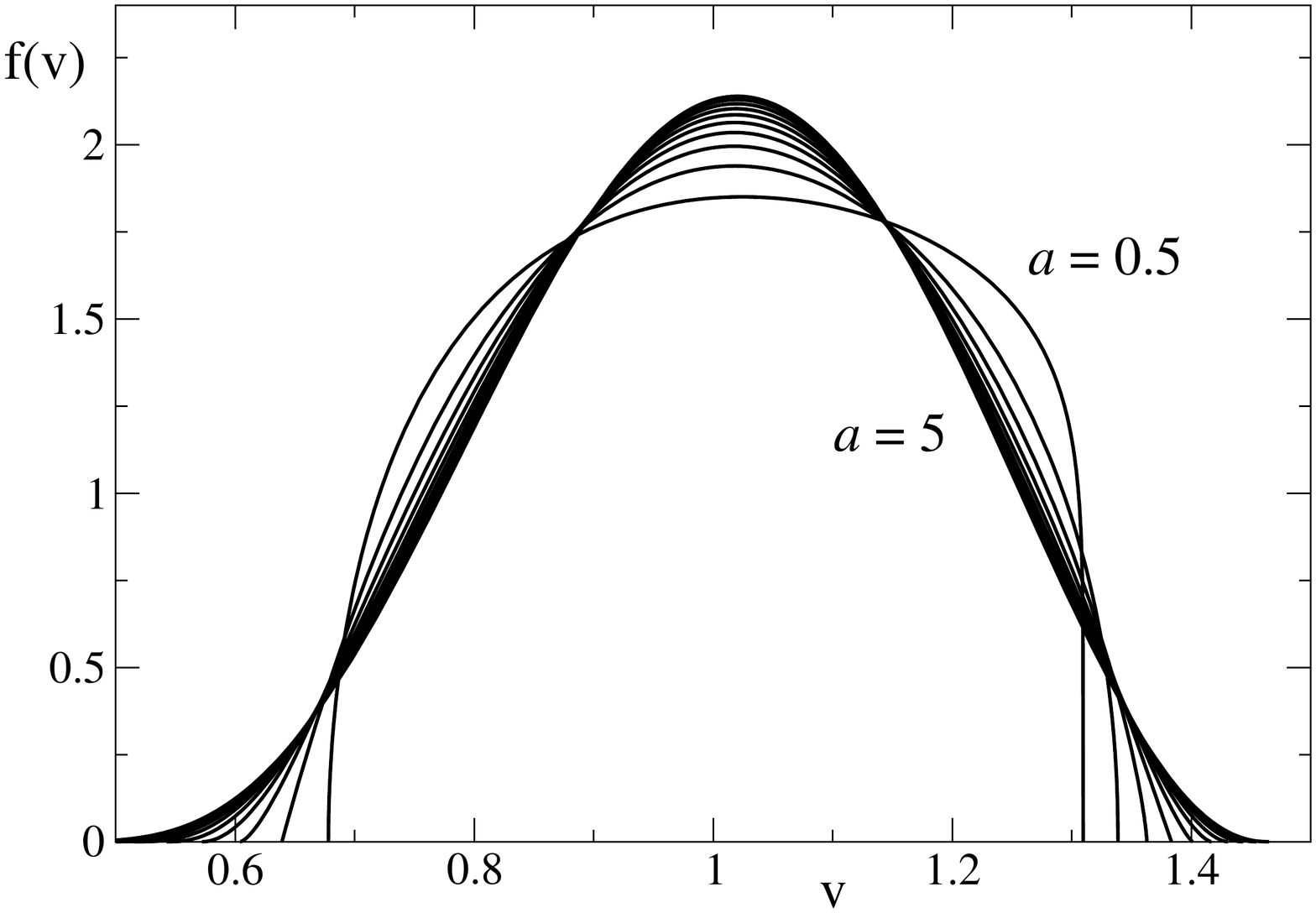}
\end{center}
  \caption{{\bf (a, top)} Examples of critical Beta size distributions,
    corresponding to the values of $(a,b)$ from
    Fig.~\protect\ref{fig:beta_a_b_plane}, with $a=0.5,1,1.5,\ldots,5$
    increasing in the direction shown. {\bf (b, bottom)} Corresponding distributions of $v=\sigma^3$, which is proportional to particle volume; these distributions are much more nearly symmetric.}
\label{fig:beta_distributions}
\end{figure}

We can now proceed as above and solve the MFE critical point criterion
to find out what shape parameters $a$ and $b$ produce critical S--SS
transitions. We do this at fixed density, taking again $n\p=1.133$. The
spinodal condition then fixes $\delta$ for given $a$ and $b$, and the
critical point gives one additional condition, so that we get a line
of critical points in the $a,b$ plane as shown in
Fig.~\ref{fig:beta_a_b_plane}. What is noticeable is that the critical
points lie significantly away from the line $a=b$ where the parent
density distribution is symmetric. This is also clearly visible in
Fig.~\ref{fig:beta_distributions}a, with the critical size
distributions having distinct peaks to the right of the mean.
One is led to ask whether there are other quantities, related but not
identical to particle diameter, that would have more symmetric
distributions. An obvious choice is the particle volume, which is
proportional to $v=\sigma^3$. As
Fig.~\ref{fig:beta_distributions}b shows, the distributions $f_v(v)
= f(\sigma)/(3\sigma^2)$ are indeed much more nearly symmetric at
criticality. This suggests that deviations from such symmetry, as
measured by the skew
\begin{equation}
s_v = \frac{\langle (v-\langle v\rangle)^3\rangle}
{\langle (v-\langle v\rangle)^2\rangle^{3/2}}
\end{equation}
indicate deviations from criticality, and conversely $s_v=0$ might be
a reasonable approximate way of identifying critical size
distributions. We have included the line in the $(a,b)$ plane that
results when we solve this condition (at the same values of $\delta$
as previously) in Fig.~\ref{fig:beta_a_b_plane}. The agreement with
the critical line calculated directly from the MFE criticality
condition is qualitatively quite good. In particular, the criterion
$s_v=0$ captures the fact that the critical size distributions are
asymmetric when expressed in terms of particle size, with $a>b$
throughout.

\begin{figure}[h]
\begin{center}
\includegraphics[type=pdf,ext=.pdf,read=.pdf,width=0.82\columnwidth,clip=true]{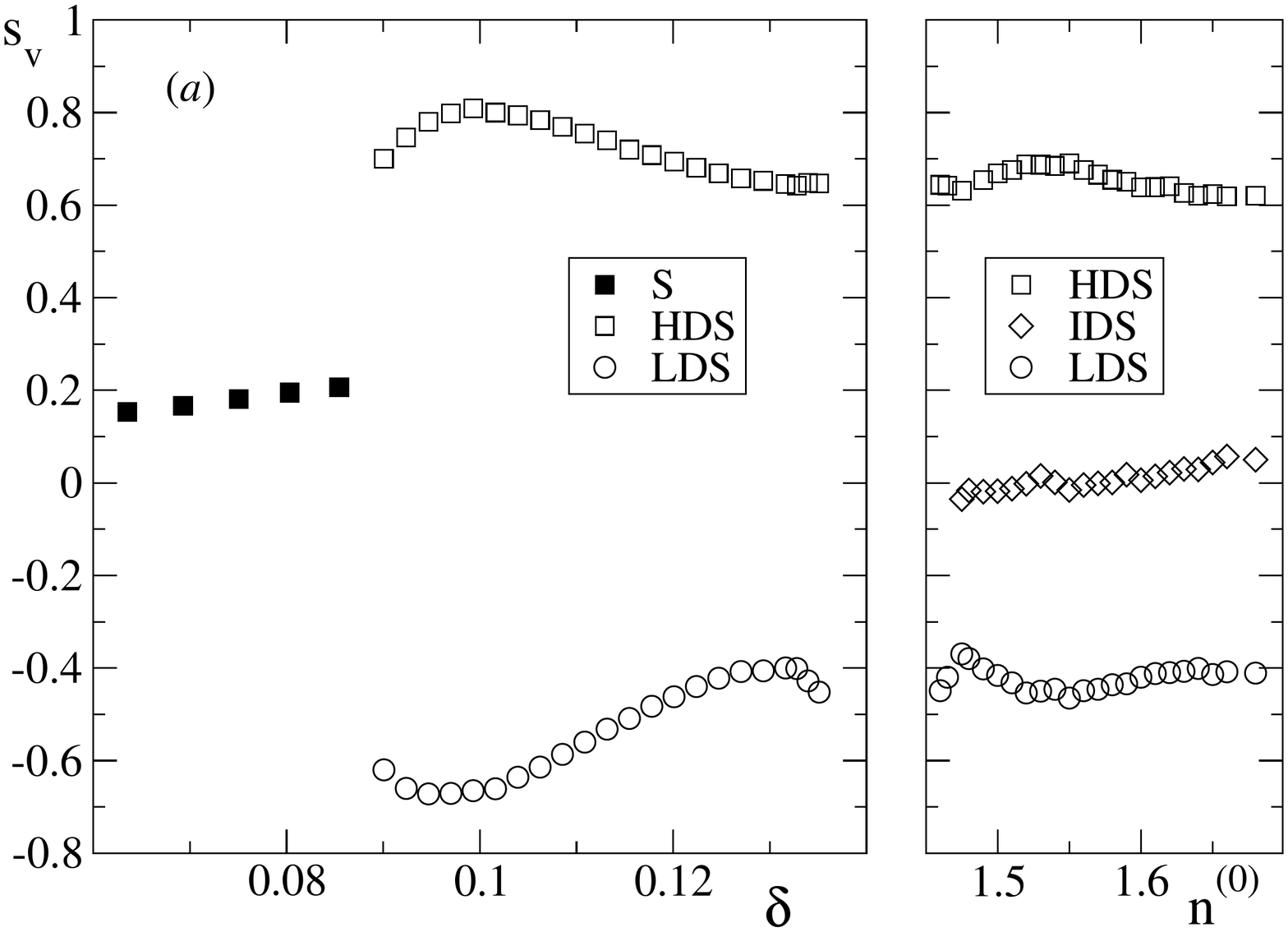}\\
\includegraphics[type=pdf,ext=.pdf,read=.pdf,width=0.82\columnwidth,clip=true]{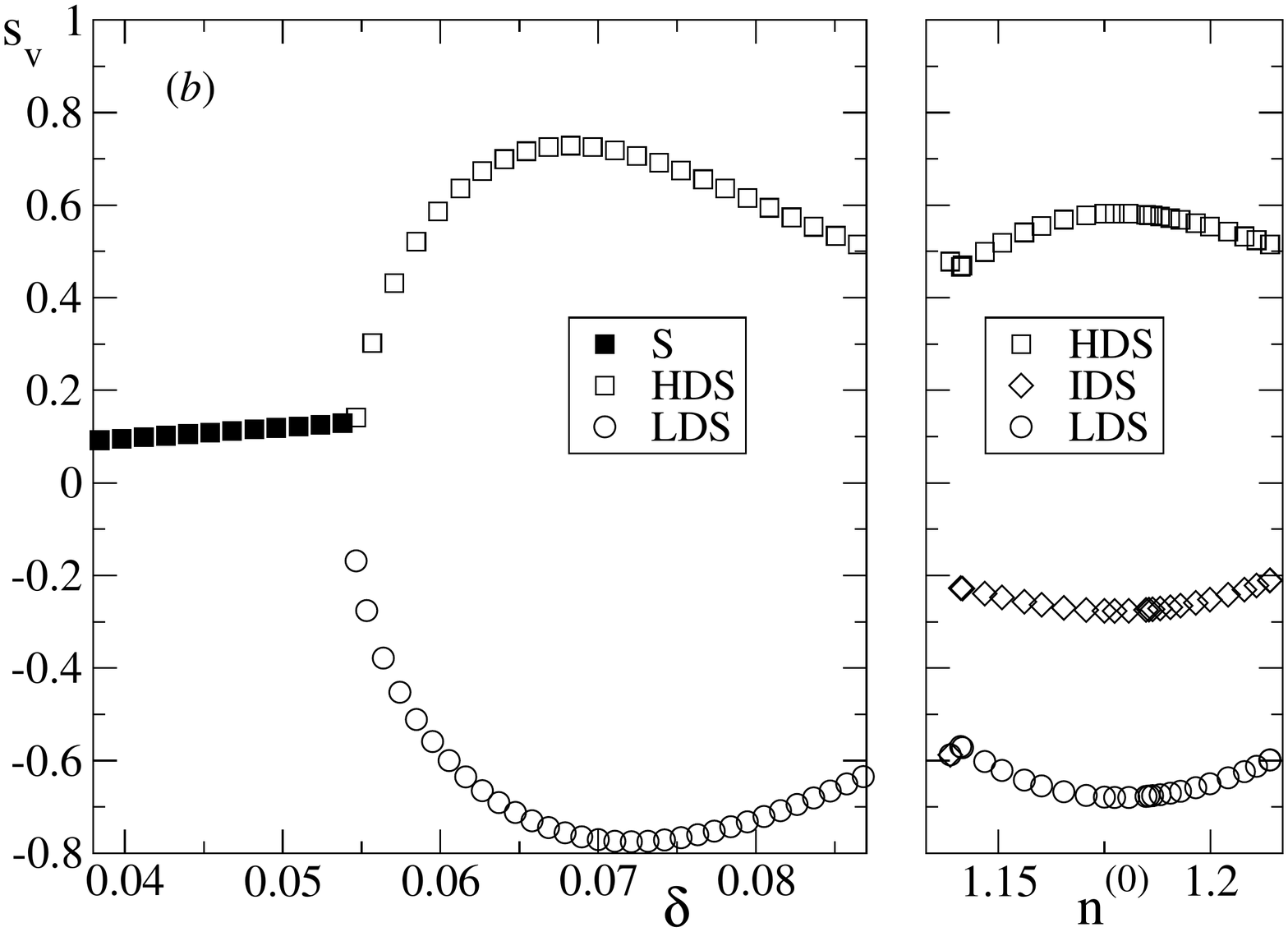}
\end{center}
\caption{Skewness $s_v$ of particle volume distribution in the phases
occurring along the phase diagram trajectories of
Fig.~\ref{fig:part_pd}. {\bf (a, top)} Simulation results, {\bf (b, bottom)} MFE calculations. Comparison with Fig.~\ref{fig:fluctuation} shows that the
near critical phases also have small $s_v$.}
\label{fig:skew3}
\end{figure}

We have also calculated $s_v$ along
the path through the phase diagram in Fig.~\ref{fig:part_pd} for top
hat parents, and show the results in
Fig.~\ref{fig:skew3}. One observes that the parent phase has
relatively low $s_v$ at the S--SS transition, in agreement with the
large values of the size fluctuation susceptibility $\chi$. Likewise, in the
SSS--SSSS transition, the phase that exhibits large $\chi$ and
splits in a near critical fashion into two solids also has small $s_v$.

Further support for the use of
$s_v=0$ as an approximate criterion for criticality comes from the
fact that $s_v$ can be written in terms of moment densities of
$\sigma$ as
\begin{eqnarray}
s_v &=& \frac{\langle v^3\rangle -3\langle v^2\rangle \langle v\rangle + 2
\langle v\rangle^3}
{(\langle v^2\rangle - \langle v\rangle^2)^{3/2}}
\\
&=& \frac{\rho_9\rho_0^2 - 3\rho_6\rho_3\rho_0 +
2\rho_3^3}
{(\rho_6\rho_0-\rho_3^2)^{3/2}}
\end{eqnarray}
which entails exactly the moment densities $\rho_0,\ldots,\rho_9$
(though not all of them) that we would expect from the general
discussion above. Nevertheless the criterion $s_v=0$ clearly remains
approximate: for the slanted top hat parents, the results in
Fig.~\ref{fig:critical_slanted} show that here the agreement with the
full criticality criterion is less good. In particular, from $s_v=0$
we would predict that there are no critical points for slant ratio
$R<1$, whereas in fact critical size distributions exist down to
$R\approx 0.81$. The question of whether there is a more accurate yet
still simple criterion for S--SS criticality remains open.

\section{Discussion and future work}
\label{sec:discussion}

In summary we have deployed tailored Monte Carlo simulation methods and
moment free energy calculations to provide conclusive evidence that
dense polydisperse spheres at equilibrium demix into coexisting fcc
phases, with more phases appearing as the spread of diameters and the
number density increase. Up to four coexisting phase were tracked, each
of which contained a narrower distribution of particle sizes than is
present in the system overall. Interestingly it was observed that for
our systems the S--SS and the SSS--SSSS transitions are quasi-critical,
characterised by a large correlation length for fluctuations in local
particle size. By contrast the SS--SSS transition was found to be
strongly first order. To rationalize these observations, we investigated
the features of the parental size distribution that control the
character of solid 
demixing transitions. It was found that small skew in the parent
distribution of particle volumes ($s_v\approx 0$) correlates well with
the existence of a quasi-continuous transition, at least for one class
of parental distribution shapes.

Whilst our results settle the matter of the true equilibrium behaviour,
they leave open the question as to the extent to which this behaviour
will be observable in experimental studies of polydisperse systems.
Initial indications from recent experiments on colloid-polymer mixtures
are that solid-solid demixing does not occur on the timescale of weeks
\cite{Poon2010}. Thus the best opportunity to see evidence may be to
focus on regions of the phase diagrams where polydisperse solid(s)
coexist with a fluid that can transport particles to their preferred
solid phase. Additionally it would be interesting to try to manufacture
a distribution of particle sizes that has $s_v\approx 0$ and then look
for an increase in particle size fluctuations in the single
solid region, even if the full transition itself is not seen. 

As regards the questions that our results pose for further simulation
and theoretical work, an interesting matter is that of the fate of the
regions of multiple solid coexistence at high volume fraction. As the
polydispersity $\delta$ is reduced, it seems
clear that all the transition lines to multiple solids (S--SS,
SS--SSS, SSS--SSSS etc) must converge on (but never quite
reach) the monodisperse close packed limit at
$\delta=0,\eta=\pi/\sqrt{18}\approx 74\%$ since the close packed crystal
will be unstable to any finite degree of polydispersity. We can also
consider what happens if we fix the polydispersity $\delta>0$ and
increase the parent volume fraction. The number of fractionated solids
will increase without bound as the pressure increases, until at some
volume fraction the pressure diverges and the system cannot be
compressed further. The locus of these points in the phase diagram forms the
infinite pressure line. Also this line must, as $\delta$ is decreased to
zero, approach the monodisperse close packed limit $\eta\approx 74\%$.

An intriguing question is whether criticality can play a role in the
approach to the close packed limit along the S--SS boundary. As we
have seen, it is primarily the parent shape that controls the nature
of the S--SS transition. Thus there may exist parent forms for which
S--SS demixing is critical at or very near to the close packed limit,
and it would be interesting to see whether a simple characterization
of such narrow critical parent size distribution forms can be found.

Further solid phases are likely to arise in the phase diagram at
values of $\delta$ beyond those that we have explored.
For example one could imagine that at very large $\delta$
(for which the system separates into multiple coexisting phases) the
smallest particles, rather than forming their own fcc phase, might
instead secrete themselves in the interstitials of the fcc solid formed
by the largest particles, thus potentially permitting the volume
fraction to exceed $74\%$.  Indeed this could be a mechanism whereby a
unimodal parental distribution might produce familiar
substitutionally-ordered phases, such as CsCl, or exotic phases such as
AB$_2$ and AB$_{13}$ that can appear in binary colloid
mixtures \cite{Schofield2005}. Investigating this question could -- in
principle -- be tackled by simulation, but is probably beyond the
present capabilities of the MFE calculations which are based
on free energies that are reliable only for small to moderate $\delta$.

{\it Acknowledgments:} Computational results were partly produced on a
machine funded by HEFCE's Strategic Research Infrastructure fund.

\appendix

\section{Comparable locations in soft and hard sphere phase
diagrams}
\label{app:comparable}

It is difficult to map from first principles the simulation phase
diagram for soft spheres to the MFE calculations for hard spheres.
Existing approaches as summarized in e.g.\  Ref.~\cite{ShuClaJon95} do
allow one to  calculate effective hard sphere diameters for soft
particles, but are based on liquid-state correlations and work only up
to moderate densities.

We therefore identified comparable points based on the phase diagram
topology. In particular, the simulations at polydispersity
$\delta=13.7\%$ show an instability towards FSS coexistence very close
to the SS--SSS transition (see Fig.~\ref{fig:part_pd}a). From the
density range between these two points, relative to the separation
between the SS--SSS and SSS--SSSS transitions, we estimate the
corresponding polydispersity for hard spheres to be $\delta=8.7\%$,
just below the meeting point of the SS--FSS and SS--SSS lines
in Fig.~\ref{fig:part_pd}b. We find the density corresponding to the
vertical trajectory in Fig.~\ref{fig:part_pd}a ($n\p=1.45$)
similarly: at this density and at $\delta=13.7\%$, the simulations
show an SS phase split that is still stable but becomes unstable at
slightly higher $\delta$. The corresponding density in the MFE phase
can be estimated as $n\p=1.133$, just below the SS--FSS transition
line at $\delta=8.7\%$. This fixes the vertical and final horizontal
trajectories through the MFE phase diagram which we use in evaluating e.g.\
the correlation volume data in Fig.~\ref{fig:fluctuation}.

For the SSSS state point in Fig.~\ref{fig:daughters4} we
proceed similarly. This point lies on the final horizontal trajectory
through the phase diagram, for which we already have the hard sphere
polydispersity $\delta=8.7\%$ that corresponds to the simulation value
$\delta=13.7\%$. We then estimate the density of the state point so
that its density difference to the SSS--SSSS transition, in units
of the separation between the SS--SSS and SSS--SSSS transitions,
is the same as in the simulations. This gave $n\p=1.232$. The same
method was applied for the SSS state point in
Fig.~\ref{fig:daughters3}. For the SS point in
Fig.~\ref{fig:daughters2}, we simply scaled the polydispersities in
proportion to the value of $\delta$ on the horizontal trajectories
through the phase diagram, so that $\delta=0.095$ in the simulations
is mapped to $\delta=0.095\times 0.087/0.137=0.0612$.

%
\bibliography{Papers,references} 

\begin{thebibliography}{50}%
\makeatletter
\providecommand \@ifxundefined [1]{%
 \@ifx{#1\undefined}
}%
\providecommand \@ifnum [1]{%
 \ifnum #1\expandafter \@firstoftwo
 \else \expandafter \@secondoftwo
 \fi
}%
\providecommand \@ifx [1]{%
 \ifx #1\expandafter \@firstoftwo
 \else \expandafter \@secondoftwo
 \fi
}%
\providecommand \natexlab [1]{#1}%
\providecommand \enquote  [1]{``#1''}%
\providecommand \bibnamefont  [1]{#1}%
\providecommand \bibfnamefont [1]{#1}%
\providecommand \citenamefont [1]{#1}%
\providecommand \href@noop [0]{\@secondoftwo}%
\providecommand \href [0]{\begingroup \@sanitize@url \@href}%
\providecommand \@href[1]{\@@startlink{#1}\@@href}%
\providecommand \@@href[1]{\endgroup#1\@@endlink}%
\providecommand \@sanitize@url [0]{\catcode `\\12\catcode `\$12\catcode
  `\&12\catcode `\#12\catcode `\^12\catcode `\_12\catcode `\%12\relax}%
\providecommand \@@startlink[1]{}%
\providecommand \@@endlink[0]{}%
\providecommand \url  [0]{\begingroup\@sanitize@url \@url }%
\providecommand \@url [1]{\endgroup\@href {#1}{\urlprefix }}%
\providecommand \urlprefix  [0]{URL }%
\providecommand \Eprint [0]{\href }%
\@ifxundefined \urlstyle {%
  \providecommand \doi  [0]{\begingroup \@sanitize@url \@doi}%
  \providecommand \@doi [1]{\endgroup \@@startlink {\doibase
  #1}doi:\discretionary {}{}{}#1\@@endlink }%
}{%
  \providecommand \doi  [0]{doi:\discretionary{}{}{}\begingroup
  \urlstyle{rm}\Url }%
}%
\providecommand \doibase [0]{http://dx.doi.org/}%
\providecommand \Doi [0]{\begingroup \@sanitize@url \@Doi }%
\providecommand \@Doi  [1]{\endgroup\@@startlink{\doibase#1}\@@Doi}%
\providecommand \@@Doi [1]{#1\@@endlink}%
\providecommand \selectlanguage [0]{\@gobble}%
\providecommand \bibinfo  [0]{\@secondoftwo}%
\providecommand \bibfield  [0]{\@secondoftwo}%
\providecommand \translation [1]{[#1]}%
\providecommand \BibitemOpen [0]{}%
\providecommand \bibitemStop [0]{}%
\providecommand \bibitemNoStop [0]{.\EOS\space}%
\providecommand \EOS [0]{\spacefactor3000\relax}%
\providecommand \BibitemShut  [1]{\csname bibitem#1\endcsname}%
\bibitem [{\citenamefont {Hales}\ and\ \citenamefont
  {Ferguson}(2006)}]{Hales2006}%
  \BibitemOpen
  \bibfield  {author} {\bibinfo {author} {\bibfnamefont {T.~C.}\ \bibnamefont
  {Hales}}\ and\ \bibinfo {author} {\bibfnamefont {S.~P.}\ \bibnamefont
  {Ferguson}},\ }\bibfield  {title} {\enquote {\bibinfo {title} {A formulation
  of the kepler conjecture},}\ }\href
  {http://dx.doi.org/10.1007/s00454-005-1211-1} {\bibfield  {journal} {\bibinfo
   {journal} {Discrete Comput. Geom.},\ }\textbf {\bibinfo {volume} {36}},\
  \bibinfo {pages} {21} (\bibinfo {year} {2006})},\ ISSN \bibinfo {issn}
  {0179-5376}\BibitemShut {NoStop}%
\bibitem [{\citenamefont {Woodcock}(1997)}]{Woodcock1997a}%
  \BibitemOpen
  \bibfield  {author} {\bibinfo {author} {\bibfnamefont {L.~V.}\ \bibnamefont
  {Woodcock}},\ }\bibfield  {title} {\enquote {\bibinfo {title} {Entropy
  difference between the face-centred cubic and hexagonal close-packed crystal
  structures},}\ }\href@noop {} {\bibfield  {journal} {\bibinfo  {journal}
  {Nature},\ }\textbf {\bibinfo {volume} {385}},\ \bibinfo {pages} {141}
  (\bibinfo {year} {1997})}\BibitemShut {NoStop}%
\bibitem [{\citenamefont {Bruce}\ \emph {et~al.}(1997)\citenamefont {Bruce},
  \citenamefont {Wilding},\ and\ \citenamefont {Ackland}}]{Bruce1997a}%
  \BibitemOpen
  \bibfield  {author} {\bibinfo {author} {\bibfnamefont {A.~D.}\ \bibnamefont
  {Bruce}}, \bibinfo {author} {\bibfnamefont {N.~B.}\ \bibnamefont {Wilding}},
  \ and\ \bibinfo {author} {\bibfnamefont {G.~J.}\ \bibnamefont {Ackland}},\
  }\bibfield  {title} {\enquote {\bibinfo {title} {Free energy of crystalline
  solids: A lattice-switch monte carlo method},}\ }\Doi
  {10.1103/PhysRevLett.79.3002} {\bibfield  {journal} {\bibinfo  {journal}
  {Phys. Rev. Lett.},\ }\textbf {\bibinfo {volume} {79}},\ \bibinfo {pages}
  {3002} (\bibinfo {year} {1997})}\BibitemShut {NoStop}%
\bibitem [{\citenamefont {Alder}\ and\ \citenamefont
  {Wainwright}(1957)}]{Alder1957}%
  \BibitemOpen
  \bibfield  {author} {\bibinfo {author} {\bibfnamefont {B.~J.}\ \bibnamefont
  {Alder}}\ and\ \bibinfo {author} {\bibfnamefont {T.~E.}\ \bibnamefont
  {Wainwright}},\ }\bibfield  {title} {\enquote {\bibinfo {title} {Phase
  transition for a hard sphere system},}\ }\Doi {10.1063/1.1743957} {\bibfield
  {journal} {\bibinfo  {journal} {J. Chem. Phys.},\ }\textbf {\bibinfo {volume}
  {27}},\ \bibinfo {pages} {1208} (\bibinfo {year} {1957})}\BibitemShut
  {NoStop}%
\bibitem [{\citenamefont {Dickinson}(1978)}]{Dickinson1978}%
  \BibitemOpen
  \bibfield  {author} {\bibinfo {author} {\bibfnamefont {E.}~\bibnamefont
  {Dickinson}},\ }\bibfield  {title} {\enquote {\bibinfo {title} {General
  discussion},}\ }\href {http://dx.doi.org/10.1039/DC9786500114} {\bibfield
  {journal} {\bibinfo  {journal} {Faraday Discuss. Chem. Soc.},\ }\textbf
  {\bibinfo {volume} {65}},\ \bibinfo {pages} {127} (\bibinfo {year}
  {1978})}\BibitemShut {NoStop}%
\bibitem [{\citenamefont {{Barrat, J.L.}}\ and\ \citenamefont {{Hansen,
  J.P.}}(1986)}]{Barrat1986}%
  \BibitemOpen
  \bibfield  {author} {\bibinfo {author} {\bibnamefont {{Barrat, J.L.}}}\ and\
  \bibinfo {author} {\bibnamefont {{Hansen, J.P.}}},\ }\bibfield  {title}
  {\enquote {\bibinfo {title} {On the stability of polydisperse colloidal
  crystals},}\ }\Doi {10.1051/jphys:019860047090154700} {\bibfield  {journal}
  {\bibinfo  {journal} {J. Phys. France},\ }\textbf {\bibinfo {volume} {47}},\
  \bibinfo {pages} {1547} (\bibinfo {year} {1986})}\BibitemShut {NoStop}%
\bibitem [{\citenamefont {Bartlett}(1998)}]{Bartlett1998}%
  \BibitemOpen
  \bibfield  {author} {\bibinfo {author} {\bibfnamefont {P.}~\bibnamefont
  {Bartlett}},\ }\bibfield  {title} {\enquote {\bibinfo {title} {Fractionated
  crystallization in a polydisperse mixture of hard spheres},}\ }\Doi
  {10.1063/1.477753} {\bibfield  {journal} {\bibinfo  {journal} {J. Chem.
  Phys.},\ }\textbf {\bibinfo {volume} {109}},\ \bibinfo {pages} {10970}
  (\bibinfo {year} {1998})}\BibitemShut {NoStop}%
\bibitem [{\citenamefont {Sear}(1998)}]{Sear1998}%
  \BibitemOpen
  \bibfield  {author} {\bibinfo {author} {\bibfnamefont {R.~P.}\ \bibnamefont
  {Sear}},\ }\bibfield  {title} {\enquote {\bibinfo {title} {Phase separation
  and crystallisation of polydisperse hard spheres},}\ }\href@noop {}
  {\bibfield  {journal} {\bibinfo  {journal} {Europhys. Lett.},\ }\textbf
  {\bibinfo {volume} {44}},\ \bibinfo {pages} {531} (\bibinfo {year}
  {1998})}\BibitemShut {NoStop}%
\bibitem [{\citenamefont {Phan}\ \emph {et~al.}(1998)\citenamefont {Phan},
  \citenamefont {Russel}, \citenamefont {Zhu},\ and\ \citenamefont
  {Chaikin}}]{Phan1998}%
  \BibitemOpen
  \bibfield  {author} {\bibinfo {author} {\bibfnamefont {S.~E.}\ \bibnamefont
  {Phan}}, \bibinfo {author} {\bibfnamefont {W.~B.}\ \bibnamefont {Russel}},
  \bibinfo {author} {\bibfnamefont {J.}~\bibnamefont {Zhu}}, \ and\ \bibinfo
  {author} {\bibfnamefont {P.~M.}\ \bibnamefont {Chaikin}},\ }\bibfield
  {title} {\enquote {\bibinfo {title} {Effects of polydispersity on hard sphere
  crystals},}\ }\Doi {10.1063/1.476453} {\bibfield  {journal} {\bibinfo
  {journal} {J. Chem. Phys.},\ }\textbf {\bibinfo {volume} {108}},\ \bibinfo
  {pages} {9789} (\bibinfo {year} {1998})}\BibitemShut {NoStop}%
\bibitem [{\citenamefont {Lacks}\ and\ \citenamefont
  {Wienhoff}(1999)}]{Lacks1999}%
  \BibitemOpen
  \bibfield  {author} {\bibinfo {author} {\bibfnamefont {D.~J.}\ \bibnamefont
  {Lacks}}\ and\ \bibinfo {author} {\bibfnamefont {J.~R.}\ \bibnamefont
  {Wienhoff}},\ }\bibfield  {title} {\enquote {\bibinfo {title} {Disappearances
  of energy minima and loss of order in polydisperse colloidal systems},}\
  }\Doi {10.1063/1.479282} {\bibfield  {journal} {\bibinfo  {journal} {J. Chem.
  Phys.},\ }\textbf {\bibinfo {volume} {111}},\ \bibinfo {pages} {398}
  (\bibinfo {year} {1999})}\BibitemShut {NoStop}%
\bibitem [{\citenamefont {Chaudhuri}\ \emph {et~al.}(2005)\citenamefont
  {Chaudhuri}, \citenamefont {Karmakar}, \citenamefont {Dasgupta},
  \citenamefont {Krishnamurthy},\ and\ \citenamefont {Sood}}]{Chaudhuri2005}%
  \BibitemOpen
  \bibfield  {author} {\bibinfo {author} {\bibfnamefont {P.}~\bibnamefont
  {Chaudhuri}}, \bibinfo {author} {\bibfnamefont {S.}~\bibnamefont {Karmakar}},
  \bibinfo {author} {\bibfnamefont {C.}~\bibnamefont {Dasgupta}}, \bibinfo
  {author} {\bibfnamefont {H.~R.}\ \bibnamefont {Krishnamurthy}}, \ and\
  \bibinfo {author} {\bibfnamefont {A.~K.}\ \bibnamefont {Sood}},\ }\bibfield
  {title} {\enquote {\bibinfo {title} {Equilibrium glassy phase in a
  polydisperse hard-sphere system},}\ }\Doi {10.1103/PhysRevLett.95.248301}
  {\bibfield  {journal} {\bibinfo  {journal} {Phys. Rev. Lett.},\ }\textbf
  {\bibinfo {volume} {95}},\ \bibinfo {pages} {248301} (\bibinfo {year}
  {2005})}\BibitemShut {NoStop}%
\bibitem [{\citenamefont {Fernandez}\ \emph {et~al.}(2007)\citenamefont
  {Fernandez}, \citenamefont {Martin-Mayor},\ and\ \citenamefont
  {Verrocchio}}]{Fernandez2007}%
  \BibitemOpen
  \bibfield  {author} {\bibinfo {author} {\bibfnamefont {L.~A.}\ \bibnamefont
  {Fernandez}}, \bibinfo {author} {\bibfnamefont {V.}~\bibnamefont
  {Martin-Mayor}}, \ and\ \bibinfo {author} {\bibfnamefont {P.}~\bibnamefont
  {Verrocchio}},\ }\bibfield  {title} {\enquote {\bibinfo {title} {Phase
  diagram of a polydisperse soft-spheres model for liquids and colloids},}\
  }\Doi {10.1103/PhysRevLett.98.085702} {\bibfield  {journal} {\bibinfo
  {journal} {Phys. Rev. Lett.},\ }\textbf {\bibinfo {volume} {98}},\ \bibinfo
  {eid} {085702} (\bibinfo {year} {2007})}\BibitemShut {NoStop}%
\bibitem [{\citenamefont {Yang}\ and\ \citenamefont {Ma}(2009)}]{Yang2009}%
  \BibitemOpen
  \bibfield  {author} {\bibinfo {author} {\bibfnamefont {M.}~\bibnamefont
  {Yang}}\ and\ \bibinfo {author} {\bibfnamefont {H.}~\bibnamefont {Ma}},\
  }\bibfield  {title} {\enquote {\bibinfo {title} {Solid-solid transition of
  the size-polydisperse hard sphere system},}\ }\Doi {10.1063/1.3056412}
  {\bibfield  {journal} {\bibinfo  {journal} {J. Chem. Phys.},\ }\textbf
  {\bibinfo {volume} {130}},\ \bibinfo {eid} {031103} (\bibinfo {year}
  {2009})}\BibitemShut {NoStop}%
\bibitem [{Note1()}]{Note1}%
  \BibitemOpen
  \bibinfo {note} {Though see reference \cite {Byelov2010} for a recent
  experimental observation of solid-solid phase separation in polydisperse
  platelike particles.}\BibitemShut {Stop}%
\bibitem [{\citenamefont {Auer}\ and\ \citenamefont
  {Frenkel}(2001)}]{Auer2001a}%
  \BibitemOpen
  \bibfield  {author} {\bibinfo {author} {\bibfnamefont {S.}~\bibnamefont
  {Auer}}\ and\ \bibinfo {author} {\bibfnamefont {D.}~\bibnamefont {Frenkel}},\
  }\bibfield  {title} {\enquote {\bibinfo {title} {Suppression of crystal
  nucleation in polydisperse colloids due to increase of the surface free
  energy},}\ }\href {http://dx.doi.org/10.1038/35099513} {\bibfield  {journal}
  {\bibinfo  {journal} {Nature},\ }\textbf {\bibinfo {volume} {413}},\ \bibinfo
  {pages} {711} (\bibinfo {year} {2001})}\BibitemShut {NoStop}%
\bibitem [{\citenamefont {Poon}(2002)}]{Poon2002}%
  \BibitemOpen
  \bibfield  {author} {\bibinfo {author} {\bibfnamefont {W.~C.~K.}\
  \bibnamefont {Poon}},\ }\bibfield  {title} {\enquote {\bibinfo {title} {The
  physics of a model colloid-polymer mixture},}\ }\href
  {http://dx.doi.org/10.1088/0953-8984/14/33/201} {\bibfield  {journal}
  {\bibinfo  {journal} {Journal of Physics: Condensed Matter},\ }\textbf
  {\bibinfo {volume} {14}},\ \bibinfo {pages} {R859} (\bibinfo {year}
  {2002})}\BibitemShut {NoStop}%
\bibitem [{\citenamefont {Zaccarelli}\ \emph {et~al.}(2009)\citenamefont
  {Zaccarelli}, \citenamefont {Valeriani}, \citenamefont {Sanz}, \citenamefont
  {Poon}, \citenamefont {Cates},\ and\ \citenamefont
  {Pusey}}]{ZacValSanPooCat09}%
  \BibitemOpen
  \bibfield  {author} {\bibinfo {author} {\bibfnamefont {E.}~\bibnamefont
  {Zaccarelli}}, \bibinfo {author} {\bibfnamefont {C.}~\bibnamefont
  {Valeriani}}, \bibinfo {author} {\bibfnamefont {E.}~\bibnamefont {Sanz}},
  \bibinfo {author} {\bibfnamefont {W.~C.~K.}\ \bibnamefont {Poon}}, \bibinfo
  {author} {\bibfnamefont {M.~E.}\ \bibnamefont {Cates}}, \ and\ \bibinfo
  {author} {\bibfnamefont {P.~N.}\ \bibnamefont {Pusey}},\ }\bibfield  {title}
  {\enquote {\bibinfo {title} {Crystallization of hard-sphere glasses},}\ }\Doi
  {10.1103/PhysRevLett.103.135704} {\bibfield  {journal} {\bibinfo  {journal}
  {Phys. Rev. Lett.},\ }\textbf {\bibinfo {volume} {103}},\ \bibinfo {pages}
  {135704} (\bibinfo {year} {2009})}\BibitemShut {NoStop}%
\bibitem [{\citenamefont {Evans}\ \emph {et~al.}(1998)\citenamefont {Evans},
  \citenamefont {Fairhurst},\ and\ \citenamefont {Poon}}]{evans1998}%
  \BibitemOpen
  \bibfield  {author} {\bibinfo {author} {\bibfnamefont {R.}~\bibnamefont
  {Evans}}, \bibinfo {author} {\bibfnamefont {D.}~\bibnamefont {Fairhurst}}, \
  and\ \bibinfo {author} {\bibfnamefont {W.}~\bibnamefont {Poon}},\ }\bibfield
  {title} {\enquote {\bibinfo {title} {Universal law of fractionation for
  slightly polydisperse systems},}\ }\href@noop {} {\bibfield  {journal}
  {\bibinfo  {journal} {Phys. Rev. Lett.},\ }\textbf {\bibinfo {volume} {81}},\
  \bibinfo {pages} {1326} (\bibinfo {year} {1998})}\BibitemShut {NoStop}%
\bibitem [{\citenamefont {Erne}\ \emph {et~al.}(2005)\citenamefont {Erne},
  \citenamefont {van~den Pol}, \citenamefont {Vroege}, \citenamefont {Visser},\
  and\ \citenamefont {Wensink}}]{Erne2005}%
  \BibitemOpen
  \bibfield  {author} {\bibinfo {author} {\bibfnamefont {B.~H.}\ \bibnamefont
  {Erne}}, \bibinfo {author} {\bibfnamefont {E.}~\bibnamefont {van~den Pol}},
  \bibinfo {author} {\bibfnamefont {G.~J.}\ \bibnamefont {Vroege}}, \bibinfo
  {author} {\bibfnamefont {T.}~\bibnamefont {Visser}}, \ and\ \bibinfo {author}
  {\bibfnamefont {H.~H.}\ \bibnamefont {Wensink}},\ }\bibfield  {title}
  {\enquote {\bibinfo {title} {Size fractionation in a phase-separated
  colloidal fluid},}\ }\href@noop {} {\bibfield  {journal} {\bibinfo  {journal}
  {Langmuir},\ }\textbf {\bibinfo {volume} {21}},\ \bibinfo {pages} {1802}
  (\bibinfo {year} {2005})}\BibitemShut {NoStop}%
\bibitem [{\citenamefont {Sollich}\ \emph
  {et~al.}(2001){\natexlab{a}}\citenamefont {Sollich}, \citenamefont {Warren},\
  and\ \citenamefont {Cates}}]{Sollich2001}%
  \BibitemOpen
  \bibfield  {author} {\bibinfo {author} {\bibfnamefont {P.}~\bibnamefont
  {Sollich}}, \bibinfo {author} {\bibfnamefont {P.~B.}\ \bibnamefont {Warren}},
  \ and\ \bibinfo {author} {\bibfnamefont {M.~E.}\ \bibnamefont {Cates}},\
  }\bibfield  {title} {\enquote {\bibinfo {title} {Moment free energies for
  polydisperse systems},}\ }\href@noop {} {\bibfield  {journal} {\bibinfo
  {journal} {Adv. Chem. Phys.},\ }\textbf {\bibinfo {volume} {116}},\ \bibinfo
  {pages} {265} (\bibinfo {year} {2001}{\natexlab{a}})}\BibitemShut {NoStop}%
\bibitem [{\citenamefont {{Wilding}}\ and\ \citenamefont
  {{Sollich}}(2010)}]{Wilding2010b}%
  \BibitemOpen
  \bibfield  {author} {\bibinfo {author} {\bibfnamefont {N.~B.}\ \bibnamefont
  {{Wilding}}}\ and\ \bibinfo {author} {\bibfnamefont {P.}~\bibnamefont
  {{Sollich}}},\ }\bibfield  {title} {\enquote {\bibinfo {title} {{Phase
  behaviour of polydisperse spheres: simulation strategies and an application
  to the freezing transition}},}\ }\href@noop {} {\bibfield  {journal}
  {\bibinfo  {journal} {ArXiv e-prints}} (\bibinfo {year} {2010})},\ \Eprint
  {http://arxiv.org/abs/1008.3068} {arXiv:1008.3068 [cond-mat.soft]}
  \BibitemShut {NoStop}%
\bibitem [{\citenamefont {Salacuse}\ and\ \citenamefont
  {Stell}(1982)}]{Salacuse1982}%
  \BibitemOpen
  \bibfield  {author} {\bibinfo {author} {\bibfnamefont {J.~J.}\ \bibnamefont
  {Salacuse}}\ and\ \bibinfo {author} {\bibfnamefont {G.}~\bibnamefont
  {Stell}},\ }\bibfield  {title} {\enquote {\bibinfo {title} {Polydisperse
  systems: Statistical thermodynamics, with applications to several models
  including hard and permeable spheres},}\ }\Doi {10.1063/1.444274} {\bibfield
  {journal} {\bibinfo  {journal} {J. Chem. Phys.},\ }\textbf {\bibinfo {volume}
  {77}},\ \bibinfo {pages} {3714} (\bibinfo {year} {1982})}\BibitemShut
  {NoStop}%
\bibitem [{\citenamefont {Fasolo}\ and\ \citenamefont
  {Sollich}(2004){\natexlab{a}}}]{Fasolo2004}%
  \BibitemOpen
  \bibfield  {author} {\bibinfo {author} {\bibfnamefont {M.}~\bibnamefont
  {Fasolo}}\ and\ \bibinfo {author} {\bibfnamefont {P.}~\bibnamefont
  {Sollich}},\ }\bibfield  {title} {\enquote {\bibinfo {title} {Fractionation
  effects in phase equilibria of polydisperse hard-sphere colloids},}\ }\Doi
  {10.1103/PhysRevE.70.041410} {\bibfield  {journal} {\bibinfo  {journal}
  {Phys. Rev. E},\ }\textbf {\bibinfo {volume} {70}},\ \bibinfo {pages}
  {041410} (\bibinfo {year} {2004}{\natexlab{a}})}\BibitemShut {NoStop}%
\bibitem [{\citenamefont {Hansen}(1970)}]{Hansen1970}%
  \BibitemOpen
  \bibfield  {author} {\bibinfo {author} {\bibfnamefont {J.~P.}\ \bibnamefont
  {Hansen}},\ }\bibfield  {title} {\enquote {\bibinfo {title} {Phase transition
  of the {Lennard}-{Jones} system. {II}. high-temperature limit},}\ }\Doi
  {10.1103/PhysRevA.2.221} {\bibfield  {journal} {\bibinfo  {journal} {Phys.
  Rev. A},\ }\textbf {\bibinfo {volume} {2}},\ \bibinfo {pages} {221} (\bibinfo
  {year} {1970})}\BibitemShut {NoStop}%
\bibitem [{\citenamefont {Hoover}\ \emph {et~al.}(1970)\citenamefont {Hoover},
  \citenamefont {Ross}, \citenamefont {Johnson}, \citenamefont {Henderson},
  \citenamefont {Barker},\ and\ \citenamefont {Brown}}]{Hoover1970}%
  \BibitemOpen
  \bibfield  {author} {\bibinfo {author} {\bibfnamefont {W.~G.}\ \bibnamefont
  {Hoover}}, \bibinfo {author} {\bibfnamefont {M.}~\bibnamefont {Ross}},
  \bibinfo {author} {\bibfnamefont {K.~W.}\ \bibnamefont {Johnson}}, \bibinfo
  {author} {\bibfnamefont {D.}~\bibnamefont {Henderson}}, \bibinfo {author}
  {\bibfnamefont {J.~A.}\ \bibnamefont {Barker}}, \ and\ \bibinfo {author}
  {\bibfnamefont {B.~C.}\ \bibnamefont {Brown}},\ }\bibfield  {title} {\enquote
  {\bibinfo {title} {Soft-sphere equation of state},}\ }\href@noop {}
  {\bibfield  {journal} {\bibinfo  {journal} {J. Chem. Phys.},\ }\textbf
  {\bibinfo {volume} {52}},\ \bibinfo {pages} {4931} (\bibinfo {year}
  {1970})}\BibitemShut {NoStop}%
\bibitem [{\citenamefont {Wilding}(2009){\natexlab{a}}}]{Wilding2009a}%
  \BibitemOpen
  \bibfield  {author} {\bibinfo {author} {\bibfnamefont {N.~B.}\ \bibnamefont
  {Wilding}},\ }\bibfield  {title} {\enquote {\bibinfo {title} {Freezing
  parameters of soft spheres},}\ }\href@noop {} {\bibfield  {journal} {\bibinfo
   {journal} {Mol. Phys.},\ }\textbf {\bibinfo {volume} {107}},\ \bibinfo
  {pages} {295} (\bibinfo {year} {2009}{\natexlab{a}})}\BibitemShut {NoStop}%
\bibitem [{\citenamefont {Sollich}(2002)}]{Sollich2002}%
  \BibitemOpen
  \bibfield  {author} {\bibinfo {author} {\bibfnamefont {P.}~\bibnamefont
  {Sollich}},\ }\bibfield  {title} {\enquote {\bibinfo {title} {Predicting
  phase equilibria in polydisperse systems},}\ }\href
  {http://stacks.iop.org/0953-8984/14/i=3/a=201} {\bibfield  {journal}
  {\bibinfo  {journal} {J.Phys: Condensed Matter},\ }\textbf {\bibinfo {volume}
  {14}},\ \bibinfo {pages} {R79} (\bibinfo {year} {2002})}\BibitemShut
  {NoStop}%
\bibitem [{\citenamefont {Sollich}\ \emph
  {et~al.}(2001){\natexlab{b}}\citenamefont {Sollich}, \citenamefont {Warren},\
  and\ \citenamefont {Cates}}]{SolWarCat01}%
  \BibitemOpen
  \bibfield  {author} {\bibinfo {author} {\bibfnamefont {P.}~\bibnamefont
  {Sollich}}, \bibinfo {author} {\bibfnamefont {P.~B.}\ \bibnamefont {Warren}},
  \ and\ \bibinfo {author} {\bibfnamefont {M.~E.}\ \bibnamefont {Cates}},\
  }\bibfield  {title} {\enquote {\bibinfo {title} {Moment free energies for
  polydisperse systems},}\ }\Doi {10.1002/9780470141762.ch4} {\bibfield
  {journal} {\bibinfo  {journal} {Adv. Chem. Phys.},\ }\textbf {\bibinfo
  {volume} {116}},\ \bibinfo {pages} {265} (\bibinfo {year}
  {2001}{\natexlab{b}})}\BibitemShut {NoStop}%
\bibitem [{\citenamefont {Warren}(1998)}]{Warren98}%
  \BibitemOpen
  \bibfield  {author} {\bibinfo {author} {\bibfnamefont {P.~B.}\ \bibnamefont
  {Warren}},\ }\bibfield  {title} {\enquote {\bibinfo {title} {Combinatorial
  entropy and the statistical mechanics of polydispersity},}\ }\Doi
  {10.1103/PhysRevLett.80.1369} {\bibfield  {journal} {\bibinfo  {journal}
  {Phys. Rev. Lett.},\ }\textbf {\bibinfo {volume} {80}},\ \bibinfo {pages}
  {1369} (\bibinfo {year} {1998})}\BibitemShut {NoStop}%
\bibitem [{\citenamefont {Sollich}\ and\ \citenamefont
  {Cates}(1998)}]{SolCat98}%
  \BibitemOpen
  \bibfield  {author} {\bibinfo {author} {\bibfnamefont {P.}~\bibnamefont
  {Sollich}}\ and\ \bibinfo {author} {\bibfnamefont {M.~E.}\ \bibnamefont
  {Cates}},\ }\bibfield  {title} {\enquote {\bibinfo {title} {Projected free
  energies for polydisperse phase equilibria},}\ }\Doi
  {10.1103/PhysRevLett.80.1365} {\bibfield  {journal} {\bibinfo  {journal}
  {Phys. Rev. Lett.},\ }\textbf {\bibinfo {volume} {80}},\ \bibinfo {pages}
  {1365} (\bibinfo {year} {1998})}\BibitemShut {NoStop}%
\bibitem [{\citenamefont {Bartlett}(1997)}]{Bartlett97}%
  \BibitemOpen
  \bibfield  {author} {\bibinfo {author} {\bibfnamefont {P.}~\bibnamefont
  {Bartlett}},\ }\bibfield  {title} {\enquote {\bibinfo {title} {A
  geometrically-based mean-field theory of polydisperse hard- sphere
  mixtures},}\ }\Doi {10.1063/1.474364} {\bibfield  {journal} {\bibinfo
  {journal} {J. Chem. Phys.},\ }\textbf {\bibinfo {volume} {107}},\ \bibinfo
  {pages} {188} (\bibinfo {year} {1997})}\BibitemShut {NoStop}%
\bibitem [{\citenamefont {Kranendonk}\ and\ \citenamefont
  {Frenkel}(1991)}]{KraFre91}%
  \BibitemOpen
  \bibfield  {author} {\bibinfo {author} {\bibfnamefont {W.~G.~T.}\
  \bibnamefont {Kranendonk}}\ and\ \bibinfo {author} {\bibfnamefont
  {D.}~\bibnamefont {Frenkel}},\ }\bibfield  {title} {\enquote {\bibinfo
  {title} {Computer-simulation of solid liquid coexistence in binary hard-
  sphere mixtures},}\ }\Doi {10.1080/00268979100100501} {\bibfield  {journal}
  {\bibinfo  {journal} {Mol. Phys.},\ }\textbf {\bibinfo {volume} {72}},\
  \bibinfo {pages} {679} (\bibinfo {year} {1991})}\BibitemShut {NoStop}%
\bibitem [{\citenamefont {Speranza}\ and\ \citenamefont
  {Sollich}(2002)}]{SpeSol02}%
  \BibitemOpen
  \bibfield  {author} {\bibinfo {author} {\bibfnamefont {A.}~\bibnamefont
  {Speranza}}\ and\ \bibinfo {author} {\bibfnamefont {P.}~\bibnamefont
  {Sollich}},\ }\bibfield  {title} {\enquote {\bibinfo {title} {Simplified
  {Onsager} theory for isotropic-nematic phase equilibria of length
  polydisperse hard rods},}\ }\Doi {10.1063/1.1499718} {\bibfield  {journal}
  {\bibinfo  {journal} {J. Chem. Phys.},\ }\textbf {\bibinfo {volume} {117}},\
  \bibinfo {pages} {5421} (\bibinfo {year} {2002})}\BibitemShut {NoStop}%
\bibitem [{\citenamefont {Speranza}\ and\ \citenamefont
  {Sollich}(2003)}]{SpeSol03a}%
  \BibitemOpen
  \bibfield  {author} {\bibinfo {author} {\bibfnamefont {A.}~\bibnamefont
  {Speranza}}\ and\ \bibinfo {author} {\bibfnamefont {P.}~\bibnamefont
  {Sollich}},\ }\bibfield  {title} {\enquote {\bibinfo {title}
  {Isotropic-nematic phase equilibria of polydisperse hard rods: the effect of
  fat tails in the length distribution},}\ }\Doi {10.1063/1.1545444} {\bibfield
   {journal} {\bibinfo  {journal} {J. Chem. Phys.},\ }\textbf {\bibinfo
  {volume} {118}},\ \bibinfo {pages} {5213} (\bibinfo {year}
  {2003})}\BibitemShut {NoStop}%
\bibitem [{\citenamefont {Kofke}\ and\ \citenamefont
  {Glandt}(1988)}]{Kofke1988}%
  \BibitemOpen
  \bibfield  {author} {\bibinfo {author} {\bibfnamefont {D.~A.}\ \bibnamefont
  {Kofke}}\ and\ \bibinfo {author} {\bibfnamefont {E.~D.}\ \bibnamefont
  {Glandt}},\ }\href@noop {} {\bibfield  {journal} {\bibinfo  {journal} {Mol.
  Phys.},\ }\textbf {\bibinfo {volume} {64}},\ \bibinfo {pages} {1105}
  (\bibinfo {year} {1988})}\BibitemShut {NoStop}%
\bibitem [{\citenamefont {Frenkel}\ and\ \citenamefont
  {Smit}(2002)}]{frenkelsmit2002}%
  \BibitemOpen
  \bibfield  {author} {\bibinfo {author} {\bibfnamefont {D.}~\bibnamefont
  {Frenkel}}\ and\ \bibinfo {author} {\bibfnamefont {B.}~\bibnamefont {Smit}},\
  }\href@noop {} {\emph {\bibinfo {title} {Understanding Molecular
  Simulation}}}\ (\bibinfo  {publisher} {Academic},\ \bibinfo {address} {San
  Diego},\ \bibinfo {year} {2002})\BibitemShut {NoStop}%
\bibitem [{\citenamefont {Wilding}(2003)}]{Wilding2003a}%
  \BibitemOpen
  \bibfield  {author} {\bibinfo {author} {\bibfnamefont {N.~B.}\ \bibnamefont
  {Wilding}},\ }\bibfield  {title} {\enquote {\bibinfo {title} {A
  nonequilibrium {Monte} {Carlo} approach to potential refinement in inverse
  problems},}\ }\href@noop {} {\bibfield  {journal} {\bibinfo  {journal} {J.
  Chem. Phys.},\ }\textbf {\bibinfo {volume} {119}},\ \bibinfo {pages} {12163}
  (\bibinfo {year} {2003})}\BibitemShut {NoStop}%
\bibitem [{\citenamefont {Buzzacchi}\ \emph {et~al.}(2006)\citenamefont
  {Buzzacchi}, \citenamefont {Sollich}, \citenamefont {Wilding},\ and\
  \citenamefont {M\"{u}ller}}]{Buzzacchi2006}%
  \BibitemOpen
  \bibfield  {author} {\bibinfo {author} {\bibfnamefont {M.}~\bibnamefont
  {Buzzacchi}}, \bibinfo {author} {\bibfnamefont {P.}~\bibnamefont {Sollich}},
  \bibinfo {author} {\bibfnamefont {N.~B.}\ \bibnamefont {Wilding}}, \ and\
  \bibinfo {author} {\bibfnamefont {M.}~\bibnamefont {M\"{u}ller}},\ }\bibfield
   {title} {\enquote {\bibinfo {title} {Simulation estimates of cloud points of
  polydisperse fluids},}\ }\Doi {10.1103/PhysRevE.73.046110} {\bibfield
  {journal} {\bibinfo  {journal} {Phys. Rev. E},\ }\textbf {\bibinfo {volume}
  {73}},\ \bibinfo {eid} {046110} (\bibinfo {year} {2006})}\BibitemShut
  {NoStop}%
\bibitem [{\citenamefont {Wilding}(2009){\natexlab{b}}}]{Wilding2009}%
  \BibitemOpen
  \bibfield  {author} {\bibinfo {author} {\bibfnamefont {N.~B.}\ \bibnamefont
  {Wilding}},\ }\bibfield  {title} {\enquote {\bibinfo {title} {Solid-liquid
  coexistence of polydisperse fluids via simulation},}\ }\Doi
  {10.1063/1.3081141} {\bibfield  {journal} {\bibinfo  {journal} {J. Chem.
  Phys.},\ }\textbf {\bibinfo {volume} {130}},\ \bibinfo {eid} {104103}
  (\bibinfo {year} {2009}{\natexlab{b}})}\BibitemShut {NoStop}%
\bibitem [{\citenamefont {Ferrenberg}\ and\ \citenamefont
  {Swendsen}(1989)}]{ferrenberg1989}%
  \BibitemOpen
  \bibfield  {author} {\bibinfo {author} {\bibfnamefont {A.~M.}\ \bibnamefont
  {Ferrenberg}}\ and\ \bibinfo {author} {\bibfnamefont {R.~H.}\ \bibnamefont
  {Swendsen}},\ }\bibfield  {title} {\enquote {\bibinfo {title} {Optimized
  {Monte}-{Carlo} data-analysis},}\ }\href@noop {} {\bibfield  {journal}
  {\bibinfo  {journal} {Phys. Rev. Lett.},\ }\textbf {\bibinfo {volume} {63}},\
  \bibinfo {pages} {1195} (\bibinfo {year} {1989})}\BibitemShut {NoStop}%
\bibitem [{\citenamefont {Borgs}\ and\ \citenamefont
  {Kotecky}(1992)}]{Borgs1992}%
  \BibitemOpen
  \bibfield  {author} {\bibinfo {author} {\bibfnamefont {C.}~\bibnamefont
  {Borgs}}\ and\ \bibinfo {author} {\bibfnamefont {R.}~\bibnamefont
  {Kotecky}},\ }\bibfield  {title} {\enquote {\bibinfo {title} {Finite-size
  effects at asymmetric 1st-order phase-transitions},}\ }\href@noop {}
  {\bibfield  {journal} {\bibinfo  {journal} {Phys. Rev. Lett.},\ }\textbf
  {\bibinfo {volume} {68}},\ \bibinfo {pages} {1734} (\bibinfo {year}
  {1992})}\BibitemShut {NoStop}%
\bibitem [{\citenamefont {Wilding}\ and\ \citenamefont
  {Sollich}(2002)}]{Wilding2002d}%
  \BibitemOpen
  \bibfield  {author} {\bibinfo {author} {\bibfnamefont {N.~B.}\ \bibnamefont
  {Wilding}}\ and\ \bibinfo {author} {\bibfnamefont {P.}~\bibnamefont
  {Sollich}},\ }\bibfield  {title} {\enquote {\bibinfo {title} {Grand canonical
  ensemble simulation studies of polydisperse fluids},}\ }\href@noop {}
  {\bibfield  {journal} {\bibinfo  {journal} {J. Chem. Phys.},\ }\textbf
  {\bibinfo {volume} {116}},\ \bibinfo {pages} {7116} (\bibinfo {year}
  {2002})}\BibitemShut {NoStop}%
\bibitem [{Note2()}]{Note2}%
  \BibitemOpen
  \bibinfo {note} {In the analogous diagram of Ref.~\cite {SolWil10}, the S--SS
  phase boundary was erroneously drawn slightly too low, at $\delta
  =8\%$.}\BibitemShut {Stop}%
\bibitem [{\citenamefont {Sollich}\ and\ \citenamefont
  {Wilding}(2010)}]{SolWil10}%
  \BibitemOpen
  \bibfield  {author} {\bibinfo {author} {\bibfnamefont {P.}~\bibnamefont
  {Sollich}}\ and\ \bibinfo {author} {\bibfnamefont {N.~B.}\ \bibnamefont
  {Wilding}},\ }\bibfield  {title} {\enquote {\bibinfo {title} {Crystalline
  phases of polydisperse spheres},}\ }\Doi {10.1103/PhysRevLett.104.118302}
  {\bibfield  {journal} {\bibinfo  {journal} {Phys. Rev. Lett.},\ }\textbf
  {\bibinfo {volume} {104}},\ \bibinfo {pages} {118302} (\bibinfo {year}
  {2010})}\BibitemShut {NoStop}%
\bibitem [{\citenamefont {Hansen}\ and\ \citenamefont
  {McDonald}(1986)}]{HanMcD86}%
  \BibitemOpen
  \bibfield  {author} {\bibinfo {author} {\bibfnamefont {J.~P.}\ \bibnamefont
  {Hansen}}\ and\ \bibinfo {author} {\bibfnamefont {I.~R.}\ \bibnamefont
  {McDonald}},\ }\href@noop {} {\emph {\bibinfo {title} {Theory of simple
  liquids (2nd ed.)}}}\ (\bibinfo  {publisher} {Academic Press},\ \bibinfo
  {address} {London},\ \bibinfo {year} {1986})\BibitemShut {NoStop}%
\bibitem [{\citenamefont {Fasolo}\ and\ \citenamefont
  {Sollich}(2004){\natexlab{b}}}]{FasSol04}%
  \BibitemOpen
  \bibfield  {author} {\bibinfo {author} {\bibfnamefont {M.}~\bibnamefont
  {Fasolo}}\ and\ \bibinfo {author} {\bibfnamefont {P.}~\bibnamefont
  {Sollich}},\ }\bibfield  {title} {\enquote {\bibinfo {title} {Fractionation
  effects in phase equilibria of polydisperse hard-sphere colloids},}\ }\Doi
  {10.1103/PhysRevE.70.041410} {\bibfield  {journal} {\bibinfo  {journal}
  {Phys. Rev. E},\ }\textbf {\bibinfo {volume} {70}},\ \bibinfo {pages}
  {041410} (\bibinfo {year} {2004}{\natexlab{b}})}\BibitemShut {NoStop}%
\bibitem [{\citenamefont {Liddle}\ \emph {et~al.}(2011)\citenamefont {Liddle},
  , \citenamefont {Narayanan},\ and\ \citenamefont {Poon}}]{Poon2010}%
  \BibitemOpen
  \bibfield  {author} {\bibinfo {author} {\bibfnamefont {S.}~\bibnamefont
  {Liddle}}, , \bibinfo {author} {\bibfnamefont {T.}~\bibnamefont {Narayanan}},
  \ and\ \bibinfo {author} {\bibfnamefont {W.}~\bibnamefont {Poon}},\ }\Doi
  {10.1021/jp037487t} {\bibfield  {journal} {\bibinfo  {journal} {J. Phys.
  Condens. Matter}} (\bibinfo {year} {2011})},\ \doi
  {10.1021/jp037487t}\BibitemShut {NoStop}%
\bibitem [{\citenamefont {Schofield}\ \emph {et~al.}(2005)\citenamefont
  {Schofield}, \citenamefont {Pusey},\ and\ \citenamefont
  {Radcliffe}}]{Schofield2005}%
  \BibitemOpen
  \bibfield  {author} {\bibinfo {author} {\bibfnamefont {A.~B.}\ \bibnamefont
  {Schofield}}, \bibinfo {author} {\bibfnamefont {P.~N.}\ \bibnamefont
  {Pusey}}, \ and\ \bibinfo {author} {\bibfnamefont {P.}~\bibnamefont
  {Radcliffe}},\ }\bibfield  {title} {\enquote {\bibinfo {title} {Stability of
  the binary colloidal crystals $ab_{2}$ and $ab_{13}$},}\ }\Doi
  {10.1103/PhysRevE.72.031407} {\bibfield  {journal} {\bibinfo  {journal}
  {Phys. Rev. E},\ }\textbf {\bibinfo {volume} {72}},\ \bibinfo {pages}
  {031407} (\bibinfo {year} {2005})}\BibitemShut {NoStop}%
\bibitem [{\citenamefont {Shumway}\ \emph {et~al.}(1995)\citenamefont
  {Shumway}, \citenamefont {Clarke},\ and\ \citenamefont
  {Jonsson}}]{ShuClaJon95}%
  \BibitemOpen
  \bibfield  {author} {\bibinfo {author} {\bibfnamefont {S.~L.}\ \bibnamefont
  {Shumway}}, \bibinfo {author} {\bibfnamefont {A.~S.}\ \bibnamefont {Clarke}},
  \ and\ \bibinfo {author} {\bibfnamefont {H.}~\bibnamefont {Jonsson}},\
  }\bibfield  {title} {\enquote {\bibinfo {title} {Molecular-dynamics
  simulations of a pressure-induced glass- transition},}\ }\Doi
  {10.1063/1.468707} {\bibfield  {journal} {\bibinfo  {journal} {J. Chem.
  Phys.},\ }\textbf {\bibinfo {volume} {102}},\ \bibinfo {pages} {1796}
  (\bibinfo {year} {1995})}\BibitemShut {NoStop}%
\bibitem [{\citenamefont {Byelov}\ \emph {et~al.}(2010)\citenamefont {Byelov},
  \citenamefont {Mourad}, \citenamefont {Snigireva}, \citenamefont {Snigirev},
  \citenamefont {Petukhov},\ and\ \citenamefont {Lekkerkerker}}]{Byelov2010}%
  \BibitemOpen
  \bibfield  {author} {\bibinfo {author} {\bibfnamefont {D.~V.}\ \bibnamefont
  {Byelov}}, \bibinfo {author} {\bibfnamefont {M.~C.~D.}\ \bibnamefont
  {Mourad}}, \bibinfo {author} {\bibfnamefont {I.}~\bibnamefont {Snigireva}},
  \bibinfo {author} {\bibfnamefont {A.}~\bibnamefont {Snigirev}}, \bibinfo
  {author} {\bibfnamefont {A.~V.}\ \bibnamefont {Petukhov}}, \ and\ \bibinfo
  {author} {\bibfnamefont {H.~N.~W.}\ \bibnamefont {Lekkerkerker}},\ }\bibfield
   {title} {\enquote {\bibinfo {title} {Experimental observation of
  fractionated crystallization in polydisperse platelike colloids},}\ }\href
  {http://dx.doi.org/10.1021/la100993k} {\bibfield  {journal} {\bibinfo
  {journal} {Langmuir},\ }\textbf {\bibinfo {volume} {26}},\ \bibinfo {pages}
  {6898} (\bibinfo {year} {2010})}\BibitemShut {NoStop}%
\end{thebibliography}%
\end{document}